\documentclass[12pt]{article}
\usepackage[dvips]{graphicx}
\usepackage{latexsym,amsmath,amsfonts,amssymb}

\newcommand{\be}{\begin{equation}}
\newcommand{\ee}{\end{equation}}
\newcommand{\beq}{\begin{equation}}
\newcommand{\eeq}{\end{equation}}
\newcommand{\bea}{\begin{eqnarray}}
\newcommand{\eea}{\end{eqnarray}}

\begin{document}
\baselineskip=15.5pt
\pagestyle{plain}
\setcounter{page}{1}


\def\del{{\partial}}
\def\vev#1{\left\langle #1 \right\rangle}
\def\cn{{\cal N}}
\def\co{{\cal O}}
\def\IC{{\mathbb C}}
\def\IR{{\mathbb R}}
\def\IZ{{\mathbb Z}}
\def\RP{{\bf RP}}
\def\CP{{\bf CP}}
\def\Poincare{{Poincar\'e }}
\def\tr{{\rm tr}}
\def\tp{{\tilde \Phi}}

\def\TL{\hfil$\displaystyle{##}$}
\def\TR{$\displaystyle{{}##}$\hfil}
\def\TC{\hfil$\displaystyle{##}$\hfil}
\def\TT{\hbox{##}}
\def\HLINE{\noalign{\vskip1\jot}\hline\noalign{\vskip1\jot}}
\def\seqalign#1#2{\vcenter{\openup1\jot
  \halign{\strut #1\cr #2 \cr}}}
\def\lbldef#1#2{\expandafter\gdef\csname #1\endcsname {#2}}
\def\eqn#1#2{\lbldef{#1}{(\ref{#1})}%
\begin{equation} #2 \label{#1} \end{equation}}
\def\eqalign#1{\vcenter{\openup1\jot
    \halign{\strut\span\TL & \span\TR\cr #1 \cr
   }}}
\def\eno#1{(\ref{#1})}
\def\href#1#2{#2}
\def\half{{1 \over 2}}

\def\ads{{\it AdS}}
\def\adsp{{\it AdS}$_{p+2}$}
\def\cft{{\it CFT}}

\newcommand{\ber}{\begin{eqnarray}}
\newcommand{\eer}{\end{eqnarray}}

\newcommand{\beqar}{\begin{eqnarray}}
\newcommand{\cN}{{\cal N}}
\newcommand{\cO}{{\cal O}}
\newcommand{\cA}{{\cal A}}
\newcommand{\cT}{{\cal T}}
\newcommand{\cF}{{\cal F}}
\newcommand{\cC}{{\cal C}}
\newcommand{\cR}{{\cal R}}
\newcommand{\cW}{{\cal W}}
\newcommand{\eeqar}{\end{eqnarray}}
\newcommand{\tht}{\thteta}
\newcommand{\lm}{\lambda}\newcommand{\Lm}{\Lambda}
\newcommand{\eps}{\epsilon}


\newcommand{\nonu}{\nonumber}
\newcommand{\oh}{\displaystyle{\frac{1}{2}}}
\newcommand{\dsl}
  {\kern.06em\hbox{\raise.15ex\hbox{$/$}\kern-.56em\hbox{$\partial$}}}
\newcommand{\id}{i\!\!\not\!\partial}
\newcommand{\as}{\not\!\! A}
\newcommand{\ps}{\not\! p}
\newcommand{\ks}{\not\! k}
\newcommand{\D}{{\cal{D}}}
\newcommand{\dv}{d^2x}
\newcommand{\Z}{{\cal Z}}
\newcommand{\N}{{\cal N}}
\newcommand{\Dsl}{\not\!\! D}
\newcommand{\Bsl}{\not\!\! B}
\newcommand{\Psl}{\not\!\! P}
\newcommand{\eeqarr}{\end{eqnarray}}
\newcommand{\ZZ}{{\rm \kern 0.275em Z \kern -0.92em Z}\;}

                                                                                                    
\def\del{{\delta^{\hbox{\sevenrm B}}}} \def\ex{{\hbox{\rm e}}}
\def\azb{A_{\bar z}} \def\az{A_z} \def\bzb{B_{\bar z}} \def\bz{B_z}
\def\czb{C_{\bar z}} \def\cz{C_z} \def\dzb{D_{\bar z}} \def\dz{D_z}
\def\im{{\hbox{\rm Im}}} \def\mod{{\hbox{\rm mod}}} \def\tr{{\hbox{\rm Tr}}}
\def\ch{{\hbox{\rm ch}}} \def\imp{{\hbox{\sevenrm Im}}}
\def\trp{{\hbox{\sevenrm Tr}}} \def\vol{{\hbox{\rm Vol}}}
\def\rl{\Lambda_{\hbox{\sevenrm R}}} \def\wl{\Lambda_{\hbox{\sevenrm W}}}
\def\fc{{\cal F}_{k+\cox}} \def\vev{vacuum expectation value}
\def\nodiv{\mid{\hbox{\hskip-7.8pt/}}}
\def\ie{{\em i.e.}}
\def\ie{\hbox{\it i.e.}}

\def\CC{{\mathchoice
{\rm C\mkern-8mu\vrule height1.45ex depth-.05ex
width.05em\mkern9mu\kern-.05em}
{\rm C\mkern-8mu\vrule height1.45ex depth-.05ex
width.05em\mkern9mu\kern-.05em}
{\rm C\mkern-8mu\vrule height1ex depth-.07ex
width.035em\mkern9mu\kern-.035em}
{\rm C\mkern-8mu\vrule height.65ex depth-.1ex
width.025em\mkern8mu\kern-.025em}}}
                                                                                                    
\def\RR{{\rm I\kern-1.6pt {\rm R}}}
\def\NN{{\rm I\!N}}
\def\ZZ{{\rm Z}\kern-3.8pt {\rm Z} \kern2pt}
\def\IB{\relax{\rm I\kern-.18em B}}
\def\ID{\relax{\rm I\kern-.18em D}}
\def\II{\relax{\rm I\kern-.18em I}}
\def\IP{\relax{\rm I\kern-.18em P}}
\newcommand{\CS}{{\scriptstyle {\rm CS}}}
\newcommand{\CSs}{{\scriptscriptstyle {\rm CS}}}
\newcommand{\rc}{\nonumber\\}
\newcommand{\bear}{\begin{eqnarray}}
\newcommand{\eear}{\end{eqnarray}}
\newcommand{\W}{{\cal W}}
\newcommand{\F}{{\cal F}}
\newcommand{\x}{{\cal O}}
\newcommand{\LL}{{\cal L}}
                                                                                                    
\def\mani{{\cal M}}
\def\calo{{\cal O}}
\def\calb{{\cal B}}
\def\calw{{\cal W}}
\def\calz{{\cal Z}}
\def\cald{{\cal D}}
\def\calc{{\cal C}}
\def\to{\rightarrow}
\def\ele{{\hbox{\sevenrm L}}}
\def\ere{{\hbox{\sevenrm R}}}
\def\zb{{\bar z}}
\def\wb{{\bar w}}
\def\nodiv{\mid{\hbox{\hskip-7.8pt/}}}
\def\menos{\hbox{\hskip-2.9pt}}
\def\dr{\dot R_}
\def\drr{\dot r_}
\def\ds{\dot s_}
\def\da{\dot A_}
\def\dga{\dot \gamma_}
\def\ga{\gamma_}
\def\dal{\dot\alpha_}
\def\al{\alpha_}
\def\cl{{closed}}
\def\cls{{closing}}
\def\vev{vacuum expectation value}
\def\tr{{\rm Tr}}
\def\to{\rightarrow}
\def\too{\longrightarrow}


\def\a{\alpha}
\def\b{\beta}
\def\c{\gamma}
\def\d{\delta}
\def\e{\epsilon}           
\def\f{\phi}               
\def\vf{\varphi}  \def\tvf{\tilde{\varphi}}
\def\vp{\varphi}
\def\g{\gamma}
\def\h{\eta}
\def\j{\psi}
\def\k{\kappa}                    
\def\l{\lambda}
\def\m{\mu}
\def\n{\nu}
\def\o{\omega}  \def\w{\omega}
\def\q{\theta}  \def\th{\theta}                  
\def\r{\rho}                                     
\def\s{\sigma}                                   
\def\t{\tau}
\def\u{\upsilon}
\def\x{\xi}
\def\z{\zeta}
\def\pt{\tilde{\varphi}}
\def\tt{\tilde{\theta}}
\def\lab{\label}  
\def\6{\partial}
\def\wg{\wedge}
\def\atanh{{\rm arctanh}}
\def\bpsi{\bar{\psi}}
\def\bt{\bar{\theta}}
\def\bvf{\bar{\varphi}}

%
                                                                                                    
\newfont{\namefont}{cmr10}
\newfont{\addfont}{cmti7 scaled 1440}
\newfont{\boldmathfont}{cmbx10}
\newfont{\headfontb}{cmbx10 scaled 1728}
\renewcommand{\theequation}{{\rm\thesection.\arabic{equation}}}
\par\hfill ITP-UU-09/12, SPIN-09/12
\par\hfill KUL-TF-09/11

\begin{center}
{\LARGE{\bf Screening effects on meson masses\\ \vskip 10pt from holography}}
\end{center}
\vskip 10pt
\begin{center}
{\large 
Francesco Bigazzi $^{a}$, Aldo L. Cotrone $^{b}$, Angel Paredes $^{c}$, \\
Alfonso V. Ramallo $^{d}$}
\end{center}
\vskip 10pt
\begin{center}
\textit{$^a$ Physique Th\'eorique et Math\'ematique and International Solvay
Institutes, Universit\'e Libre de Bruxelles; CP 231, B-1050
Bruxelles, Belgium.}\\
\textit{$^b$  Institute for theoretical physics, K.U. Leuven;
Celestijnenlaan 200D, B-3001 Leuven,
Belgium.}\\
\textit{$^c$ Institute for Theoretical Physics, Utrecht University; Leuvenlaan 4,
3584 CE Utrecht, The Netherlands.
}\\
\textit{$^d$ Departamento de F\'\i sica de Part\'\i culas, 
Universidade de Santiago de Compostela and Instituto Galego de F\'\i sica de Altas Enerx\'\i as (IGFAE);
 E-15782, Santiago de Compostela, Spain.}\\
{\small fbigazzi@ulb.ac.be, Aldo.Cotrone@fys.kuleuven.be, A.ParedesGalan@uu.nl, alfonso@fpaxp1.usc.es}
\end{center}

\vspace{15pt}

\begin{center}
\textbf{Abstract}
\end{center}

\vspace{4pt}{\small \noindent 
We study the spectra of scalar and vector mesons in four dimensional strongly coupled SQCD-like theories 
in the Veneziano limit. The gauge theories describe the low energy dynamics of intersecting
D3 and D7-branes on the singular and deformed conifold and their strong coupling regime can be 
explored by means of dual fully backreacted supergravity backgrounds. The mesons we focus on 
are dual to fluctuations of the worldvolume gauge field on a probe D7-brane in these backgrounds. 
As we will comment in detail, the general occurrence of various UV pathologies in the D3-D7 set-ups 
under study, forces us to adapt the standard holographic recipes  to theories with intrinsic cutoffs.
Just as for QED, the low energy spectra for mesonic-like bound states will  be consistent
and largely independent of the UV cutoffs. We will study in detail how these spectra vary with the number 
of the fundamental sea flavors and their mass.
}

\vfill

\newpage
\section{Introduction}

Holography is nowadays a standard and powerful method to investigate properties of some strongly coupled gauge theories.
Most of the studies of flavor physics in this context have focused on the quenched approximation where the internal quark loops are neglected.
The extension of these investigations to the unquenched cases, where the full dynamics of the flavors is included, is of obvious interest.
The main focus of this paper is the study of mass spectra of low spin ($J=0,1$) mesons in certain strongly coupled SQCD-like theories in the unquenched Veneziano regime, where the number of colors and the number of flavors are both taken to be very large, with their ratio taken to be fixed. The aim is to extract the dependence of the spectra on the number, $N_f$, of sea flavors and on their mass, $m_q$.

The ${\cal N}=1$ SQCD-like models we will consider describe the low energy dynamics at the 4d intersection of ``color'' D3-branes and $N_f$ homogeneously smeared ``flavor'' D7-branes on the singular and on the deformed conifold. The models are the flavored unquenched versions of the conformal Klebanov-Witten (KW) \cite{Klebanov:1998hh} and the confining Klebanov-Strassler (KS) \cite{ks} ones. The strong 't Hooft coupling regime of these theories is mapped to some dual backgrounds, arising as the full backreaction of the color and the flavor D-branes, with $N_f$ D7-brane sources. The supergravity background dual to the KW (resp. KS) model coupled to chiral (resp. non chiral) massless dynamical flavors was found in \cite{Benini:2006hh} (resp. \cite{Benini:2007gx}). The solution with chiral massless flavors in the KS case, which we will not consider, appears in \cite{benini}.
These massless-flavored solutions have generically (good) singularities at the origin of the transverse radial coordinate. The singularity is avoided in the massive case. The supergravity dual of the KW model coupled to chiral (resp. non chiral) dynamical massive flavors was found in \cite{Bigazzi:2008zt} (resp. \cite{Bigazzi:2008ie}). The relevant solution in the KS case with non chiral massive flavors was found in \cite{Bigazzi:2008qq}.

In order to extract the mesonic spectra we will probe these backgrounds with an external D7-brane and  study the fluctuations of a selected set of decoupled modes (dual to scalar and vector mesons) of the gauge field on its worldvolume. This study will require a careful treatment of the boundary conditions of the fluctuating fields, in connection with the fact that the backreacted D3-D7 backgrounds - unlike the corresponding unflavored ones - have various UV pathologies. 
In particular, we will show that there is a singularity in the holographic $a$-function at finite radial position.
Just as in QED, which has a UV Landau pole, the spectra of  the various bound states will nevertheless be meaningful and independent of the cutoffs (up to corrections suppressed by the UV scale). 

Since the goal of this paper is to study the dependence of the mesonic spectra on the number of sea flavors and their masses, we will have to decide how to interpret our results, i.e. how to compare different theories with different flavor parameters. As already noticed in \cite{Bigazzi:2008qq}, there is no obvious natural scale or coupling which can be assumed to stay fixed when varying those parameters. In any case, the comparison between two theories will necessarily require us to make a choice of what to keep fixed. The comparison will strongly depend on this choice. To disentangle the possible different conclusions we will need some explicit formula for the mesonic masses as a function of the physical parameters in the theory.

We will parameterize the meson masses as $M_{meson} = \Lambda_Q \,\omega$ where $\Lambda_Q$ is a dimensionful scale which
can be ``measured'' in the IR and which we will take fixed when comparing different theories. This scale will be related to the ``coupling'' and the effective ``string tension'' at the probe quark mass scale. The coefficients $\omega$ will then be determined from a numerical analysis. 

Our results show that the $\omega$'s receive small ``radiative'' corrections due to dynamical flavor loops and decrease as the effective coupling $g_{FT}^2 N_f$ (where $g_{FT}^2\sim e^{\phi}$) is increased (or if the mass of the sea flavors is decreased). 
These small ``quantum effect'' corrections resemble  the hydrogen atom Lamb shift in QED. In section \ref{rough} we will argue that, for the flavored KW models, these corrections could be related to the small variation of the effective number of (adjoint plus bifundamental) degrees of freedom due to the internal flavor loops. In the KS cases they could be related to corrections to the glueball (or Kaluza Klein) mass scale.
The decrease of the $\omega$'s depends on the choice of keeping fixed the coupling at the probe quark mass: had we kept fixed the coupling at some larger UV scale, larger $g_{FT}^2 N_f$ could have yielded larger meson masses.

The structure of the paper is the following. We will start, in section \ref{gene}, by examining the general UV behavior of the D3-D7 models under study. We will show how non trivial UV cutoffs emerge when the holographic $a$-function is considered. Moreover, we will outline the general limits  of our analysis. We will then focus on scalar and vector mesons in the flavored conifold models. In section \ref{sec:KW}, combining the numerical shooting technique and the WKB analysis, we will extract the spectra of mesons having massive constituent flavor fields in the (flavored) KW models. Having chosen a prescription to compare different theories we will study how the sea flavors (either massless or massive) affect the mesonic spectra. In section \ref{sec: KS} we will present the results of analogous studies in the flavored KS models, limiting our analysis to the case of mesons with massless fundamental constituents. In section \ref{rough} we will discuss a possible way to interpret our results. We will end in section \ref{conclu} with a set of concluding remarks. Various technical details and some comments on high spin mesons will be left to the appendices.
\section{General comments}
\label{gene}
\setcounter{equation}{0}
A common feature of D3-D7 set-ups, both in the case of smeared and localized \cite{localized} D7-branes, is a running dilaton in the dual supergravity solutions. The dilaton, and consequently the effective string coupling, typically increases with the transverse radial variable $u$ of the backgrounds and blows up at a certain point $u_d$. The solutions can thus be defined at best only in a region $u\le u_{cut}<u_{d}$ where the dilaton can stay small.\footnote{Just to fix the notation consistently with the following sections, we will call $\rho$ (resp $\tau$) the radial variable in the flavored KW (resp. KS) solutions and we will set $u_{d}=\rho_{LP}$ (resp. $u_{d}=\tau_d$). In our conventions large values of $u$ are mapped to large energy scales in the dual field theory.} 

The point $u_{d}$ can be formally mapped to a UV Landau-pole in the dual field theories. The simplest way to realize it is to consider a localized D3-D7 set-up in flat space. The theory on $N_c$ D3-branes is just ${\cal N}=4$ SYM, which has exactly zero beta function. We can add $N_f$ flavors to the theory by means of a stack  of D7-branes, sharing with the D3-branes the 4d Minkowski directions and extended along a non compact 4-dimensional submanifold of the transverse space. This breaks supersymmetry (to ${\cal N}=2$) as well as conformal invariance. The perturbatively exact coefficient of the beta function for the inverse squared gauge coupling is $b_0= 3N_c - 3N_c -N_f = -N_f$, which means that the theory has a UV Landau pole. This pathological UV behavior is inherited by D3-D7 set-ups both in orbifold and conifold models.

Apart from the dilaton divergence, the theories we are going to consider have other less explicit UV singularities.  In the flavored KW cases the integration constants can be fixed so that the warp factor vanishes exactly at the Landau pole $u_d$. In the flavored KS cases this is not a consistent choice and the warp factor will diverge badly (going to minus infinity) at $u_{d}$. This means that there will be a point $u_h<u_{d}$ where the warp factor vanishes and the metric becomes singular. A possible interpretation of this point is in terms of a duality wall in the dual theory \cite{Benini:2007gx}. Since for $u> u_h$ the warp factor becomes negative, $u_h$ replaces $u_{d}$ as a sensible UV ``end'' of the flavored KS backgrounds. 

Both the flavored KW and KS models have an even more ``hidden'' potential UV pathology. It shows up by considering their 5d reduction. Starting with a 10d string frame metric of the form
\be
ds^2 = \alpha(u)\left[dx_{\mu} dx^{\mu} + \beta(u) du^2\right] + ds_{int}^2\,,
\ee
the reduction on the internal five dimensional manifold (with volume $V_{int}(u)$), gives a 5d Einstein frame metric of the form
\be
ds_5^2 = H(u)^{1/3}\left[dx_{\mu} dx^{\mu} + \beta(u) du^2\right]\,,
\ee
where
\be
H(u)=e^{-4\phi(u)}V_{int}(u)^2 \alpha(u)^3 \,.
\label{Hdef}
\ee
In standard set-ups, the function $H(u)^{1/6}$, which can be roughly identified with the dual field theory energy scale, monotonically varies with the radial coordinate. This is also required in order for the ``holographic $a$-function'' \cite{holaf} 
\be
a(u)\sim \beta(u)^{3/2} H(u)^{7/2} [H'(u)]^{-3}\,,
\label{holadef}
\ee
to be finite.\footnote{The monotonicity of $H(u)$ also plays a crucial role in holographic computations of the entanglement entropy, see \cite{kkm}. The notations of that paper are used in the equations above.} 

\begin{figure}
 \centering
\includegraphics[width=.4\textwidth]{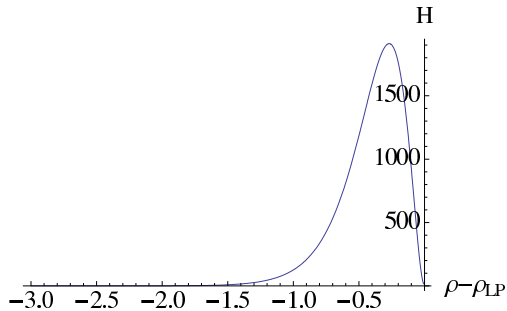}
\caption{ The function $H$ in the massless-flavored KW model.}
\label{Hflav}
\end{figure}

In both the flavored KW and KS cases, instead, the function $H(u)$ is not monotonic: it increases with $u$ from zero up to a maximum
 at a point $u_a$ and then it decreases back to zero (at $u_{d}$ in the KW cases and at $u_h$, with $u_a<u_h<u_{d}$ in the KS cases). A representative plot is given in figure \ref{Hflav}. 
 This behavior implies that the holographic $a$-function $a(u)$ as defined in (\ref{holadef}) is singular and discontinuous at $u_a$. As we will show in the following,
the $a$-function will be positive and increasing up to $u=u_a^-$ where it will asymptotically diverge. At $u=u_a^+$ $a(u)$ will go to minus infinity and then it will increase to zero (at $u_{d}$ in the KW case and at $u_h$ in the KS one) for increasing $u$. In order to avoid the apparently pathological region $u_a\le u\le u_{d}$, we will cutoff our D3-D7 backgrounds at $u_{cut} < u_a$. 
 
The occurrence of UV pathologies in the D3-D7 models we are going to consider, does not subtract interest to the study of their IR dynamics.
QED, the most notable quantum field theory model with a UV Landau pole, certainly provides excellent and meaningful predictions on the low energy (e.g. atomic) spectra, which do not sensitively  depend on the UV cutoff. In order to study, say, the spectrum of bound states of (s)quarks pairs in our D3-D7 conifold models we will just have to reconsider the standard holographic recipes and adapt them to the case where the dual supergravity backgrounds do not have a boundary at infinity. Having discussed the presence of UV pathologies, we stress that
the aim of this paper is not to study how to cure them or how to consistently UV complete the theory.
Conversely, we focus in computing IR quantities that are only mildly affected by whatever the UV physics is,
in a sense that we will make precise below.

In the following we will focus on vector and scalar mesons dual to fluctuations
 of the gauge field on the worldvolume of a D7-brane probe. The  corresponding mass spectrum will be quantized after imposing that  the fluctuations be regular at $u_Q$ (the minimal radial distance reached by the brane) and {\it vanish} at the cutoff $u_{cut}<u_a$. The pathological UV region will thus  be  excluded and treated as producing an infinite wall in the effective quantum mechanical description of the fluctuating modes. The choice of the cutoff (but not of its maximal value) is nevertheless arbitrary in the range $u_Q\ll u_{cut}< u_a$ and the consistency of our results will be guaranteed provided we show that the dependence on the cutoff is highly suppressed.
Note that we can think about our models as the IR regions of UV consistent theories, providing some UV completions to the backgrounds at hand.
This perspective has been employed recently  in  \cite{Mia:2009wj}.

Let us conclude with the following remarks. The theories to which we add flavors 
correspond to the low energy dynamics of $N$ regular (and $M$ fractional) D3-branes on the (deformed) conifold.
They are (cascading) ${\cal N}=1$ 4d gauge theories with gauge group $SU(N)\times SU(N)$ ($SU(N+M)\times SU(N)$) and bifundamental matter fields $A, B$ transforming as $SU(2)\times SU(2)$ doublets and interacting with a quartic superpotential $W_{KW}=\epsilon^{ij}\epsilon^{kl}A_iB_kA_jB_l$.
The perturbative superpotential in the flavored case is:
\begin{equation}
W = W_{KW} +\hat h_1\, \tilde q_1 (A_1B_1-A_2B_2 )q_1 + \hat h_2\,\tilde q_2 (B_1A_1-B_2A_2)q_2 + k_i\,(\tilde q_i q_i)^2 + m\,(\tilde q_i q_i)\,\,,
\label{superpot}
\end{equation}
where $q_i$ and $\tilde q_i$ are the flavor multiplets and $\hat h_i$ and $k_i$ are the couplings. 

The SQCD-like models under study 
are assumed to be in the Veneziano regime. This means that we take $N_c, N_f\rightarrow\infty$ (where $N_c$ is the number of colors) with $N_f/N_c$ and $\lambda=g_{FT}^2 N_c$ fixed. Moreover, the dual supergravity solutions (with DBI+WZ source terms for the smeared D7-branes) will only be reliable provided $\lambda\gg1$ (as usual) 
and $N_f\ll N_c$.\footnote{It is worth pointing out that the $N_f\ll N_c$ condition, necessary in all the D3-D7 models,
is not mandatory  in different set-ups with 
other kinds of  brane intersections. Some examples where $N_f/N_c$ can
be kept of order one have been studied in \cite{tutti0,tutti,tutti1}.} These limits will allow us to consider cases where the effective coupling $g_{FT}^2 N_f = N_f\lambda/N_c$, weighting the vacuum polarization effects due to the dynamical flavors, is of order one.\footnote{Note that the open string coupling on the flavor branes is {\emph{not}} $N_f e^{\phi}$: the backreacting branes are smeared in the transverse space, so only a small fraction of them is within a distance of order $\sqrt{\alpha'}$ and the coupling is parametrically smaller than $N_f e^{\phi}$ \cite{Bigazzi:2008zt,tutti1}.}

\section{Meson excitations in the KW models with flavor}
\label{sec:KW}
\setcounter{equation}{0}

In this section we will discuss (part of) the
spectrum of meson masses in the particular framework of the
so-called Klebanov-Witten conformal model \cite{Klebanov:1998hh} and
its generalizations with massless \cite{Benini:2006hh}
and massive \cite{Bigazzi:2008ie} unquenched flavors.
We start by writing general expressions for the background and excitations
and then specialize  the study for the different cases.
We will close the discussion of the flavored KW models by analyzing
the holographic $a$-function in
section \ref{sec:holakw}.


\subsection{The background solution and the excitation equations}

We consider solutions of type IIB supergravity coupled to a
homogeneously smeared set of $N_f$ D7-branes (we refer
the reader to \cite{Benini:2006hh,Bigazzi:2008zt,Bigazzi:2008ie} for
further details). The ansatz for the fields which take non-trivial
values in the solution is, in Einstein frame:
\bear
ds_{10}^2 &=& h^{-\frac12} dx_{1,3}^2 + \alpha' h^\frac12\,
\left[e^{2f}d\rho^2 + ds_5^2\right]\,\,,\rc
ds_5^2 &=& \frac{e^{2g}}{6} \sum_{i=1,2} (d\theta_i^2 +
 \sin^2 \theta_i d\varphi_i^2)+\frac{e^{2f}}{9}(d\psi+
\sum_{i=1,2} \cos\theta_i d\varphi_i)^2\,\,,\rc
F_{(5)}&=& d^4x \wedge d(h^{-1}) -    
\frac{\pi\,g_s\,N_c\,\alpha'^2}{4}\sin\theta_1 \sin\theta_2
d\theta_1 \wedge d\varphi_1 \wedge d\theta_2 \wedge d\varphi_2 \wedge d\psi\,\,, \rc
\phi&=&\phi(\rho)\,\,,\rc
F_{(1)}&=&\frac{g_s\,N_f(\rho)}{4\pi}(d\psi +\cos\theta_1 d\varphi_1
+\cos \theta_2 d\varphi_2)\,\,.
\label{ansatzgeneral}
\eear
Notice that the 5-form is self-dual  and, using the
relations:
\be
\frac{1}{2\kappa_{(10)}^2}=\frac{1}{(2\pi)^7 g_s^2 \alpha'^4}\,\,,\qquad
\qquad
T_p = \frac{1}{g_s (2\pi)^p \alpha'^{\frac{p+1}{2}}}\,,
\ee
 one can check that it satisfies the
quantization condition
$\int_{X_5}F_{(5)}= N_c T_3 2\kappa_{(10)}^2=(2\pi)^4\alpha'^2
g_s N_c$, where $X_5$ is the internal manifold. 
If the two-form  $\Omega$ is the density distribution of the  smeared D7-branes 
($g_s\Omega=-dF_{(1)}$), the action of the gravity+branes system reads \cite{Benini:2006hh}:
\bear
S&=&\frac{1}{2\kappa_{10}^2}\int d^{10}x \sqrt{-G}\left[
R -\frac12 \partial_M \phi \partial^M \phi -\frac12 e^{2\phi} |F_{(1)}|^2
-\frac14 |F_{(5)}|^2\right] +\rc
&&-T_7\left[\int d^{10}x\ e^\phi\sqrt{-G} |\Omega |
+\int C_8 \wedge \Omega \right]\,\,.
\eear
Then (\ref{ansatzgeneral}) provides a supersymmetric solution 
if:
\bear
\dot g &=& e^{2f-2g}\,\,,\qquad\qquad\quad
\dot f=3-2 e^{2f-2g} - \frac{3 g_s N_f(\rho)}{8\pi} e^\phi\,\,,\rc
\dot \phi &=& \frac{3 g_s N_f(\rho)}{4\pi} e^\phi \,\,, \qquad\qquad
\dot h = -27 \pi g_s N_c e^{-4g}\,\,.
\label{firsordeqs}
\eear
These equations are valid for the case without unquenched 
flavors \cite{Klebanov:1998hh} 
($N_f(\rho)=0$ such that $g=f=\rho$), the case with massless
unquenched flavors \cite{Benini:2006hh} ($N_f(\rho) =const >0$) and
the case with massive unquenched flavors \cite{Benini:2006hh,Bigazzi:2008zt,Bigazzi:2008ie}
(where $N_f(\rho)$ becomes a suitable
$\rho$-dependent expression depending on the kind of flavor branes under
consideration). 

In the following, we analyze excitations of 
a brane probe which preserves  the same supersymmetry
as the background. Thus, there are two kinds of flavor branes in the
generic set-up:
the first kind corresponds to  dynamical quarks, {\it i.e.} the $N_f$
branes backreacting on the geometry, accordingly corresponding to unquenched
flavors (in section \ref{sec:KWmassive}
we will denote $\rho_q$ the value of $\rho$ at their
tip).  The second kind is a single oscillating probe brane
(with tip at $\rho=\rho_Q$), associated to the
quarks which actually constitute the mesons under consideration. Of course it is possible to take
this single fluctuating brane to be one out of the $N_f$ backreacting ones but
in the following we deal with the more generic case of allowing 
 non-equal masses for dynamical and
test quarks ($\rho_q \neq \rho_Q$). Since the flavor branes affect the values of the
 metric, dilaton and p-forms
in the solutions, one may wonder whether it makes sense to consider the 
oscillation of a brane in a background where the closed string fields are taken to
be constant. That is indeed the case: a meson is associated to the oscillation of one
(or a pair of) flavor brane and not to a collective oscillation of a set of order $N_f$
flavor branes. Thus, the possible backreaction of the mesonic oscillation on the closed
string background  strictly vanishes in the Veneziano limit.

Let us consider a  flavor brane probe of the type first discussed
in \cite{kuper} and corresponding to the embedding $z_4-z_3 = e^{\frac32 \rho_Q}$,\footnote{An alternative embedding has been considered in \cite{Ouyang:2003df}.} where $z_{3,4}$ are two of the complex coordinates defining the conifold.
Details and notations are 
spelled out in appendix \ref{app: A1}.
We will only study a reduced subset of all possible mesonic modes.
Concretely, we will not discuss fluctuations of the embedding and only the
following fluctuations
of the worldvolume gauge field, which correspond to a vector and a scalar in the
dual gauge theory:
\be
{\cal A} =  a_v(\r) e^{i kx} \xi_\m dx^\m+
a_s(\r) e^{i kx} h_1\,\,.
\label{vectorans}
\ee
Here, $\xi_\mu$ is a constant transverse vector and $h_1$ is the (angular) left-invariant one-form defined in (\ref{hdefs}). 
We have chosen this particular set of fluctuations for the sake of 
simplicity, since, as we will see below, it gives rise to relatively simple
decoupled differential equations both in the present set-up and in the
Klebanov-Strassler case to be discussed in section \ref{sec: KS}.
Also for simplicity, we have 
not included any angular dependence for the fluctuating gauge field
(a truncation which is non-trivially consistent). Even if we restrict
ourselves to this very limited subset of all the possible excitations,
we expect that the results can point out the general trends on how the
presence of dynamical flavors affects the meson masses.

The procedure to obtain the second order equations associated
to these modes is outlined in appendix \ref{app: A2}. One gets:
\bear
0&=&\partial_\rho\left(e^{2g-3\r}(e^{3\r}-e^{3\r_Q})\partial_\r a_v \right)
+ M_v^2 \alpha' h\, e^{2g+2f}
\left(1+ e^{3\r_Q - 3\r} (\frac34 e^{2g-2f} -1) \right) a_v\,\,,
\label{aeq}\\
0&=&\partial_\rho\left(\frac{1-e^{3\r_Q-3\r}}{h}\partial_\r a_s \right)
-\frac32 \partial_\r (h^{-1})\,a_s-
\frac{9}{4h(1-e^{3\r_Q-3\r})}a_s+\rc
&&+M_s^2 \alpha' \, e^{2f}
\left(1+ e^{3\r_Q - 3\r}(\frac32 e^{2g-2f} -1) \right) a_s\,\,,
\label{beq}
\eear
where $M_{v,s} = - k^2$ are the  masses for the vector and scalar
mesonic excitations under consideration. 
As it is customary, enforcing appropriate IR and UV behaviors for the functions
$a_v(\rho),a_s(\rho)$ will select a discrete spectrum for $M_{v,s}$.

Before turning to the study of this issue in several cases, let us introduce
a useful parameterization of the meson masses, factoring out a dimensionful scale. Define:
\beq
M_{v,s} = \sqrt{2\pi} \left(\frac{32}{27}\right)^\frac14
 \frac{T_{Q}^\frac12}{\lambda_Q^\frac14}\ \omega_{v,s}\,\,,
 \label{M-omega}
\eeq
where $T_Q$ is the tension of a hypothetical fundamental string
 stretched
at constant $\rho = \rho_Q$ and $\lambda_Q$ is the `t Hooft 
coupling
\footnote{
\label{footcoupling}
The value of the tension $T_Q$ of eq. (\ref{TQdefs}) can be obtained by studying the short distance behavior of the $\bar Q Q$ potential, obtained by analyzing a hanging open string in the 
background metric, see appendix \ref{highspin}. 
Notice that $T_Q$ is simply the fundamental string tension $1/(2\pi\alpha')$ redshifted by the string frame warp factor. 

For the identification with the `t Hooft 
coupling, we have used the
orbifold relation
$\frac{4\pi}{g_1^2}+\frac{4\pi}{g_2^2}=g_s^{-1}e^{-\phi}$.
Let us consider $g_1^2 = g_2^2 \equiv g_{FT}^2 = 8\pi g_s e^\phi$, and 
thus define the 't Hooft coupling as
$\lambda \equiv N_c g_{FT}^2 = 8 \pi g_s N_c e^\phi $. One should keep in mind that
this orbifold relation is not guaranteed to hold in general. The relations
(\ref{TQdefs}) should be taken as the definition of $T_Q$, $\lambda_Q$.}
at the same scale $\rho = \rho_Q$:
\be
T_{Q}=\frac{1}{2\pi \a'} 
\left(e^{\phi/2} \sqrt{G_{tt}G_{xx}}\right)|_{\rho=\rho_Q}\,\,,\qquad\qquad
\lambda_Q \equiv 8\pi g_s N_c e^\phi|_{\rho=\rho_Q}\,\,.
\label{TQdefs}
\ee
The important point is that these are quantities ``measured'' in the
IR, {\it i.e.} at the tip of the brane $\rho=\rho_Q$, which is related to the
quark mass. The $\omega_{v,s}$ will be towers of numbers which, as we 
will see, depend on $\frac{N_f}{N_c}\lambda_Q$.

Notice, however, that, when writing (\ref{M-omega}), we could have 
chosen to factor out a different
IR dimensionful scale. Basically, this is related to the observation in \cite{Peet:1998wn}
of the possibility of different definitions of the radius-energy relation
in non-conformal theories.
For instance, we could have factored out the constituent quark mass
to be discussed in section \ref{sec: const}. 
Thus, one should keep in mind that
when we compute how $\omega$ varies with $\frac{N_f}{N_c}\lambda_Q$, 
we will be computing how meson masses change while keeping  
$T_{Q}^\frac12/\lambda_Q^\frac14$ fixed. 

\subsection{Mesons in the quenched approximation}

We want to analyze the equations (\ref{aeq}), (\ref{beq}) when the
background is just the unflavored KW solution, {\it i.e.}:
\be
N_f(\rho)=0\,\,,\qquad\quad 
f=g=\rho\,\,,\qquad\quad  
\phi = const
\,\,,\qquad\quad
h=\frac{27}{4}\pi g_s N_c e^{-4\rho}\,\,.
\label{unflavoredKW}
\ee
It is useful to define:
\be
\bar \rho = \rho - \rho_Q\,\,.
\label{bardef}
\ee
The equations (\ref{aeq}), (\ref{beq}) then read:
\bear
0&=& \partial_{\bar\rho}(e^{-{\bar\rho}}(e^{3{\bar\rho}} - 1) 
\partial_{\bar\rho} a_v({\bar\rho}) )
+ 
\omega_v^2 (1 - \frac{e^{-3{\bar\rho}}}{4}) a_v({\bar\rho}) \,\,,
\label{nonchiveceq}
\\
0&=&\partial_{\bar\rho}(e^{{\bar\rho}}(e^{3{\bar\rho}} - 1) 
\partial_{\bar\rho} a_s({\bar\rho}) )-
e^{4\bar\rho} \frac{33e^{3\bar\rho}-24}{4(e^{3\bar\rho}-1)}a_s({\bar\rho})
+ 
\omega_s^2
e^{2\bar\rho}(1 + \frac{e^{-3{\bar\rho}}}{2}) a_s({\bar\rho}) \,\,,
\label{nonchira1eq}
\eear
where we have used the definitions (\ref{M-omega}) and (\ref{TQdefs}).
In this conformal case, the actual quark bare mass coincides
with the
constituent mass defined as the energy of a string stretched from the
bottom of the geometry up to the brane tip $\rho_Q$:
\be
m_Q = \frac{1}{2\pi\alpha'}\int_{-\infty}^{\rho_Q}
 \sqrt{\alpha'} e^\frac{\phi}{2}e^f d\rho=\frac{e^\frac{\phi}{2}e^{\rho_Q}}
{2\pi\sqrt{\alpha'}}\,\,.
\label{quenchedmq}
\ee
This can be used to rephrase  (\ref{M-omega}) as:
\be
M_{v,s}=\frac{2\pi m_Q}{\lambda_Q^\frac12}\ \sqrt{\frac{32}{27}}\,\omega_{v,s}\,\,.
\label{Mvs}
\ee
Comparing (\ref{M-omega}) to (\ref{Mvs}), we see that
$T_Q \sim m_Q^2 /\sqrt{\lambda_Q}$.

We can now numerically  analyze equations (\ref{nonchiveceq}),
(\ref{nonchira1eq}). In each case, requiring regularity at $\bar\rho=0$
and normalizability in the UV selects a discrete set of values
for the $\omega$'s. A standard computation using the shooting technique yields:
\bear
\omega_v = 2.337 ,\ 4.720 ,\ 7.088,\ 9.454,\ \dots \rc
\omega_s =5.174,\ 7.358 ,\ 9.482,\ 11.580,\ \dots
\label{Kwmasses}
\eear

\subsubsection{The Schr\"odinger potential formalism and WKB estimates}
\label{sec: schro}

\begin{figure}[!ht]
\centering
\includegraphics[width=0.4\textwidth]{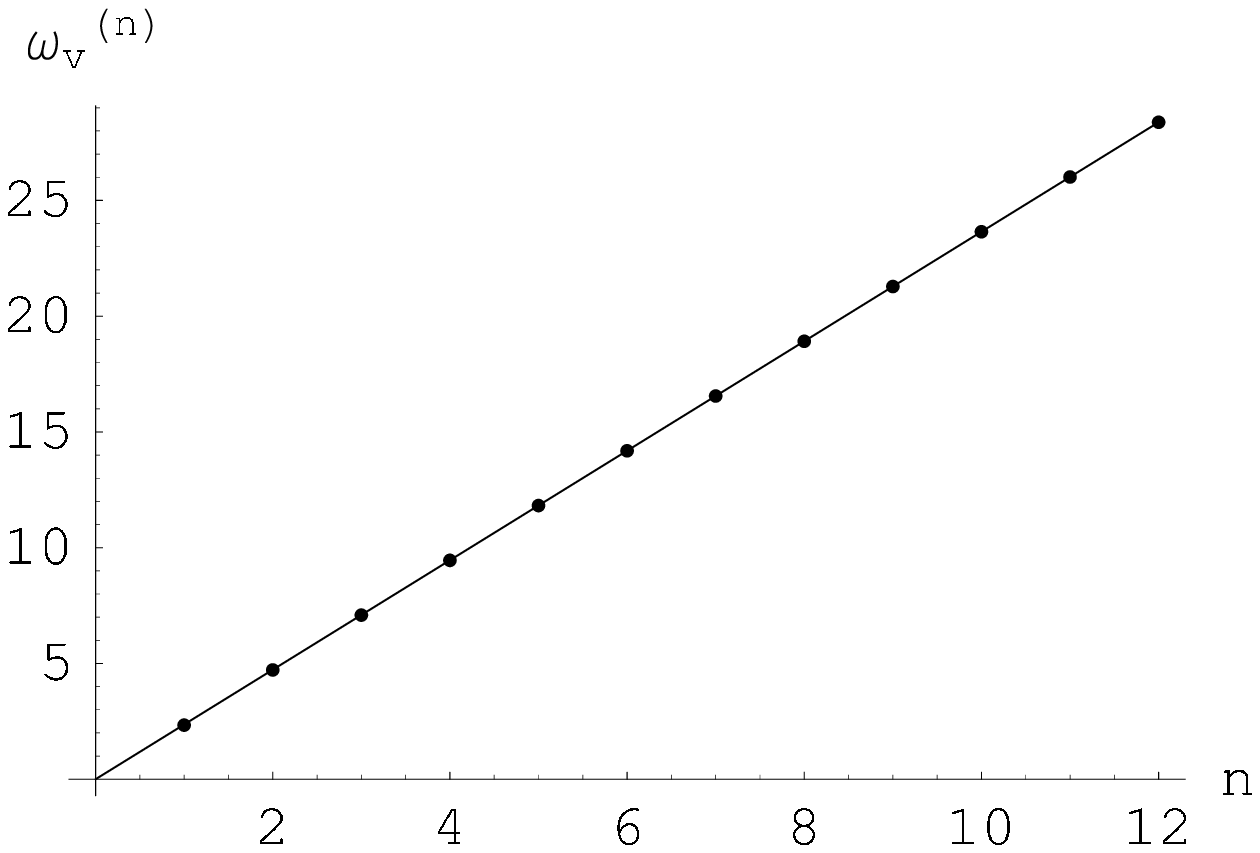} 
\includegraphics[width=0.4\textwidth]{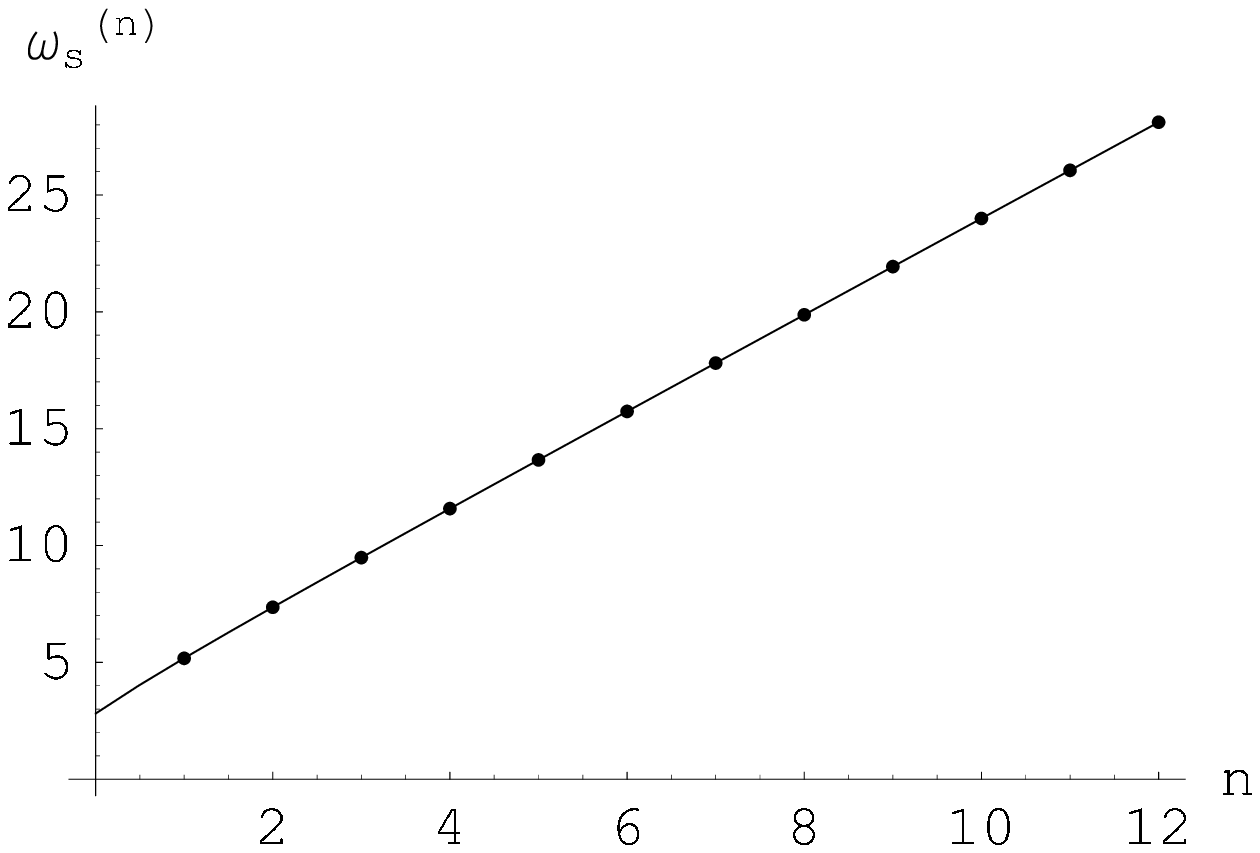} 
\caption{A plot of the first few values of $\omega_v^{(n)}$ and
$\omega_s^{(n)}$ obtained numerically (dots) and the approximation
given in eq (\ref{WKBKW}) (solid line). }
\label{KWfig}
\end{figure}

As shown in appendix \ref{schr-appendix}, after performing a  convenient change of variables, the fluctuations equations (\ref{nonchiveceq}) and (\ref{nonchira1eq}) can be written as Schr\"odinger equations for some particular potentials. We can then apply the WKB approximation to get an estimate of the mass levels. In the case of the fluctuations described by  eqs.  (\ref{nonchiveceq}) and (\ref{nonchira1eq}) this analysis leads to the following estimates of $\omega_{v,s}$:
\be
\omega_v^{(n)}\,  \approx \z_v n \,\,,\qquad\quad
\omega_s^{(n)} \approx \z_s \sqrt{n^2 + \frac72 n + \frac{15}{8}}\,\,,\qquad
\qquad (n=1,2,\dots)
\label{WKBKW}
\ee
where $\z_v$ and $\z_s$ are given by:
\be
\z_v\approx 2.365  \,\,,\qquad\quad
\z_s \approx 2.051 \,\,.
\label{WKBKWquen}
\ee
In fact, from WKB we only get terms of order $n^1$ and $n^0$ as 
$n\to \infty$. Thus, the term $\frac{15}{8}$ inside the square root
of the 
expression for $\omega_s$ does not come from WKB and we obtained it by
fitting the numerical data. In figure \ref{KWfig}, we compare these
expressions to the first few numerically found eigenvalues.


\subsection{Mesons in the presence of unquenched massless flavors}
\label{sec: masslessKW}

In this section, the background defined by the equations in
(\ref{firsordeqs})
with constant $N_f(\rho)=N_f>0$ \cite{Benini:2006hh} will be considered. The relevant
solution is:
\bear
e^\phi &=& \frac{4\pi}{3g_s N_f (\rho_{LP}-\rho)}\,\,,\rc
e^f &=& c_3 \sqrt{6(\rho_{LP}-\rho)}\, 
(1+6 (\rho_{LP}-\rho))^{-\frac13}
e^{(\rho-\r_{LP})}\,\,,
\rc
e^g &=&c_3 (1+6 (\rho_{LP}-\rho))^\frac16 e^{(\rho-\r_{LP})} 
 \,,
\rc
h&=&27 \pi g_s N_c \frac{1}{2c_3^4}
\left(\frac{1}{18e^2}
\right)^\frac13 \left(\Gamma\left(\frac13,-\frac23-4(\rho_{LP}-\rho)
\right) -\Gamma\left(\frac13,-\frac23\right) \right)\,\,,
\label{c10expr}
\eear
where the $\Gamma$ represents the incomplete gamma-function.

It is worth commenting on the integration
constants appearing in 
(\ref{c10expr}). As compared to \cite{Benini:2006hh}, we
have explicitly kept the integration constant $\rho_{LP}$ which fixes the
position of the Landau pole (where the dilaton diverges). 
We have also kept the constant $c_3$ which can
 be reabsorbed 
by rescaling the Minkowski coordinates and thus just rescales what one
defines as energy. On the other
hand, we have set the constant $c_1$ (see
eqs. (2.37), (2.38) of \cite{Benini:2006hh}) to zero
for the sake
of IR regularity. Even if the unquenched solution with
massless flavors is always IR singular, it was
shown in \cite{Benini:2006hh} that $c_1=0$ produces the less severe
IR singularity. Most importantly, it was explicitly shown in
\cite{Bigazzi:2008zt,Bigazzi:2008ie} that by introducing any non-zero
mass for the dynamical quarks, a regular IR can be obtained. Taking the
massless limit of this family of massive regular solutions
 yields the $c_1=0$ condition.
Finally, we have set $h(\rho_{LP})=0$, such that metric and dilaton
are both singular at the same point $\rho=\rho_{LP}$.  
This UV prescription is not important for the computation of the meson masses:
enforcing, instead, $h(\rho_h)=0$ for some
$\rho_h < \rho_{LP}$ would just add an
additive constant to $h$. As long as $\rho_h \gg \rho_Q$, this would only
modify the values of the meson masses by quantities exponentially
suppressed as $e^{\rho_Q - \rho_h}$. Extended comments regarding the effects
of UV prescriptions
on the computation of the masses  can be found in sections
\ref{sec: masslessSchro} and \ref{sec:EMSW}.

By inserting (\ref{c10expr}) into (\ref{aeq}), (\ref{beq}), and using again
the definitions (\ref{M-omega}), (\ref{TQdefs}), one finds the appropriate
 second order equations. It is easy to check that the resulting spectrum
of $\omega$'s
only depends on $\rho_{LP}-\rho_Q$, which can be related to the physical
quantity
$\frac{N_f}{N_c}\lambda_Q$ by using the explicit solution for the dilaton in
(\ref{c10expr}):
\be
\bar\rho_{LP}=\rho_{LP}-\rho_Q=  \frac{32\pi^2 N_c}{3N_f \lambda_Q }\,\,.
\label{rhoLP}
\ee
The
quenched case is recovered when $\frac{N_f}{N_c}\lambda_Q \to 0$, or
$\rho_{LP}-\rho_Q \to \infty$.


\subsubsection{Schr\"odinger potentials and UV cutoffs}
\label{sec: masslessSchro}

Our goal now is to study equations (\ref{aeq}), (\ref{beq}) in this 
background.
In each equation we have to demand regularity at $\rho_Q$ for the 
function that defines the fluctuation. On the other hand,
one cannot use the usual UV normalizability condition since there is
a Landau pole which severely modifies the UV behavior. 
As anticipated in section \ref{gene},
the natural
condition is to require for each excitation that:
\be
a_v(\rho_{cut})=a_s(\rho_{cut})=0\,\,.
\label{UVcutff}
\ee
These conditions amount to 
introducing by hand an infinite wall at $\rho_{cut}$. 
Our working prescription will be to set $\rho_{cut} \lesssim \rho_a$, {\it i.e.}
we consider a UV completion of the theory slightly below the scale in which the holographic
$a$-function becomes singular.

We now turn to
 justifying the proviso (\ref{UVcutff}).
 Schr\"odinger potentials associated with the
fluctuation equations can be computed following the definitions of appendix
\ref{schr-appendix}.
Figure \ref{unquenschro} shows two examples of Schr\"odinger potentials for
the vector excitation. For the scalar excitation, the plots are  qualitatively
similar (except that $V$ diverges towards $+\infty$ in the UV pathological region).

\begin{figure}[!ht]
\centering
\includegraphics[width=0.4\textwidth]{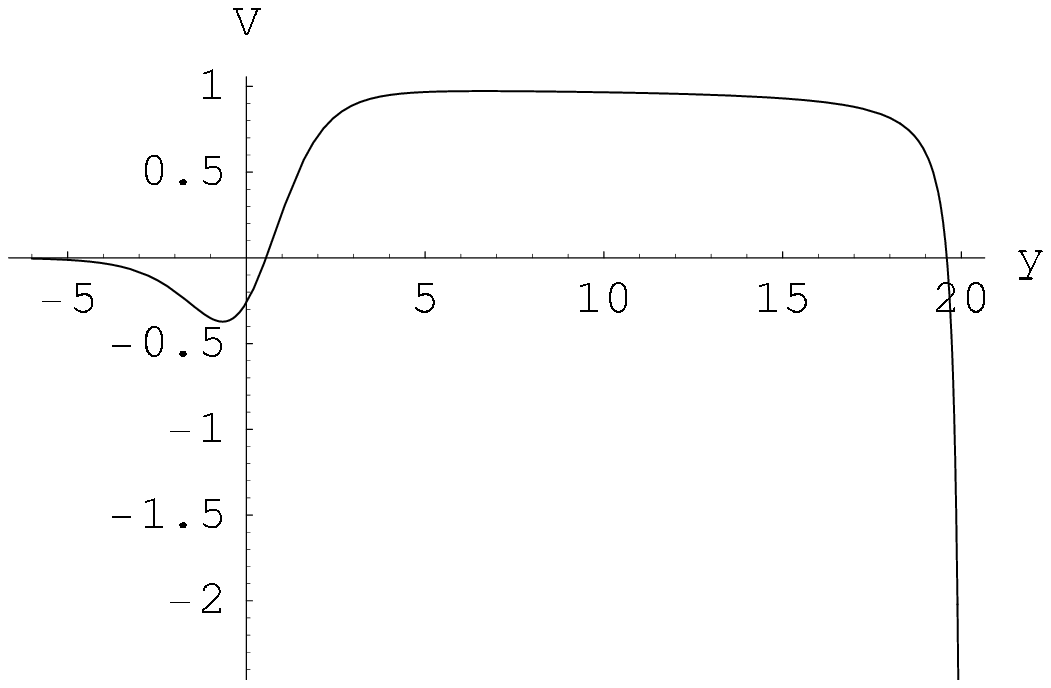} 
\includegraphics[width=0.4\textwidth]{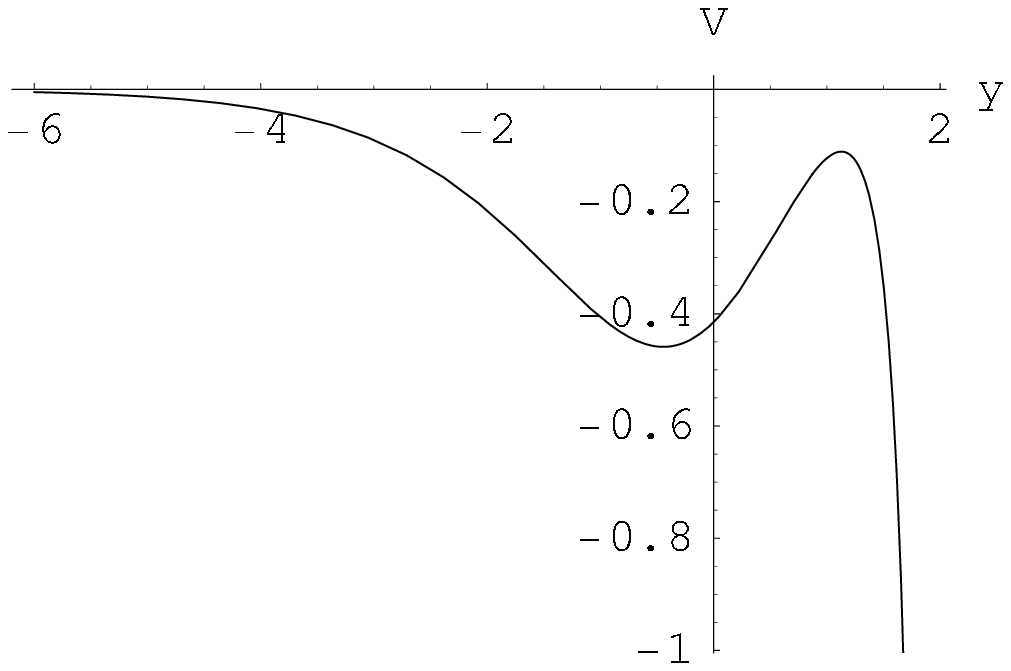} 
\caption{On the left, the Schr\"odinger potential for the vector excitation
with $\omega_v =3$, $\bar\rho_{LP}  = 20$. The UV pathological region is well separated
from the IR region.
On the right, the potential for $\omega_v =3$, $\bar\rho_{LP}  = 2$. Obviously,
the UV region is not separated from the IR one. The potentials diverge to 
minus infinity at $y=\log (e^{\bar\rho_{LP}} -1)$. }
\label{unquenschro}
\end{figure}

 From the figure, it is apparent that as long as 
$\bar \rho_{LP}$ is large ($\frac{N_f}{ N_c}\lambda_Q$ is small),
the UV pathological region lies
very far from the minimum of the potential where the physical wave function
has its main support (see the plot on the left). 
 On general grounds, we expect
that whatever would be the UV completion of the potential, it
 only affects the eigenvalues by exponentially suppressed quantities.
In particular, the precise value of $\rho_{cut}<\rho_a < \rho_{LP}$ has only a
negligible effect on the spectrum, as long as $\rho_{cut}$ lies far away
from the IR region where the potential reaches its minimum.

On the other hand, if $\frac{N_f}{ N_c}\lambda_Q$ is too large, $\rho_{LP}$
(and thus $\rho_a, \rho_{cut}$)
is not far from the IR region. Then, the formalism breaks 
down. In physical terms, if the UV completion sets in not far from the relevant IR
scale, it will affect the IR physics in a non-negligible way.
The  plot on the right of figure \ref{unquenschro} shows an example of this behavior.
For this reason, we will restrict ourselves to 
$\frac{N_f}{ N_c}\lambda_Q < 10$.

\subsubsection{Estimates of the meson spectrum from WKB}
\label{sec:EMSW}

We have checked that the WKB expressions (\ref{WKBKW}) 
in the unquenched case are also  in very good agreement with
the numerical values obtained from the shooting 
technique.\footnote{The relative error of the (\ref{WKBKW}) formula with respect to the
obtained numerical values is always well below $1\%$, except for
$\omega_{v}^{(1)}$, when in the quenched case the error is already
$1.2\%$ and in the unquenched cases remains approximately of the same order.
The same applies to the case of unquenched massive flavors to be studied
in the next section.} Thus, we  just
concentrate on studying the WKB integrals (\ref{sigmaV}), (\ref{sigmaS}).

For small $N_f \lambda_Q/N_c$ (large $\rho_{LP}-\rho_Q$),
 we can expand the integrands in (\ref{sigmaV}), (\ref{sigmaS}).
A simple computation shows that:
\bear
\z_v &\approx& 2.365(1 - 1.12\times 10^{-3} \frac{N_f}{ N_c}\lambda_Q + \dots)\,\,,\rc
\z_s &\approx& 2.051 (1- 1.08\times 10^{-3} \frac{N_f}{ N_c}\lambda_Q + \dots)\,\,.
\label{lamb}
\eear
It is clear from (\ref{lamb}) that the leading correction to the masses, due to the 
dynamical flavors, is small.
It is also approximately linear in $N_f \lambda_Q/N_c
=N_f g_{FT}^2|_{\rho=\rho_Q}$, see the bottom lines in figure \ref{fig: massive1}.  The fact that the meson masses decrease with $N_f$ might appear counter-intuitive at first sight. In fact, there is nothing strange about it,
as we will argue in the subsection \ref{physint}.
Moreover, a qualitatively similar behavior can be inferred from the unquenched lattice calculation reported in \cite{davies}. 

Now that we have found (\ref{lamb}),
let us be more precise about the size of the uncertainties of the spectrum associated
to UV prescriptions. For large $\bar\rho$ (but $\bar\rho<\bar\rho_{LP}$), the integrand 
of the WKB integrals (\ref{sigmaV}), (\ref{sigmaS}) behaves as $e^{-\bar\rho}$. Thus,
if one takes a different value of $\rho_{cut}$, the variation of the integral
and thus of the masses is of order $e^{-\bar\rho_{cut}}$. Taking into account the
radius energy relation $e^\rho \sim \Lambda$ (where $\Lambda$ is the energy scale),
one sees that different UV completions yield variations of the meson masses of order
$\frac{\Lambda_{IR}}{\Lambda_{UV}}$, where the IR scale is here associated to the
mass of the quark constituents of the meson and $\Lambda_{UV}$ the scale at which
the UV completion sets in. Notice that these corrections are far more suppressed than
the first corrections displayed in (\ref{lamb}), which are of order 
$(\bar\rho_{LP})^{-1}$ (logarithmic in the energy scale).

\subsubsection{Comments on the physical interpretation of the results}
\label{physint}

From eqs. (\ref{M-omega}), (\ref{WKBKW}), (\ref{lamb}),
we read the following expression for the tower of
vector mesons:
\be
M_{v}^{(n)} \approx 6.19\, n\, \frac{T_Q^\frac12}{\lambda_Q^\frac14}
(1 - 1.12\times 10^{-3} \frac{N_f}{ N_c}\lambda_Q + \dots)\,\,.
\label{fullformula}
\ee
From this formula, it seems that meson masses decrease with $N_f$, what,
in turn, seems to contradict the fact that dynamical quarks screen the
color charges: the binding energy between quark and anti-quark should decrease
with increasing $N_f$ and, accordingly, meson masses should increase.
The crucial point is that
the flavor effects on the meson masses heavily depend on the scales we keep fixed while comparing 
different theories. 
This means that (\ref{fullformula}) states that meson masses decrease with $N_f$
{\it {if the IR quantities $T_Q$ and $\lambda_Q$ are kept fixed when comparing theories with different
$N_f$}.} Had we kept fixed the gauge coupling at some UV scale, larger $N_f$
would have yielded smaller $\lambda_Q$, possibly resulting in larger meson masses. We now clarify
this statement with an example.

Suppose we want to rewrite (\ref{fullformula}) in terms of $T_Q$ and of the 't Hooft coupling
$\lambda_*$
at a scale $\rho_* > \rho_Q$. Due to the Landau pole, it does not make sense to
take $\rho_*$ in the far UV, and in fact let us assume $\rho_* \ll \rho_{LP}$.
Using (\ref{TQdefs}), (\ref{c10expr}), (\ref{rhoLP}), we find that:
\be
\lambda_Q = \lambda_* \left( 1 - (\rho_* - \rho_Q) \frac{3 N_f \lambda_*}{32 \pi^2 N_c} + \dots
\right)\,\,.
\ee
Equation (\ref{fullformula}) can be rewritten as:
\be
M_{v}^{(n)} \approx 6.19\, n\, \frac{T_Q^\frac12}{\lambda_*^\frac14}
\left(1 + 1.12\times 10^{-3} \frac{N_f}{ N_c}\lambda_*\Big(-1 + 2.12 (\rho_* - \rho_Q)\Big) + \dots\right)\,\,.
\label{fullformulab}
\ee
Thus, when we compare theories with different $N_f$ keeping fixed $T_Q$ and the coupling at
any scale $\rho_* > \rho_Q + 0.47 $, masses do indeed increase with $N_f$.

Analogous considerations apply to the scalar mesons and to the rest of the cases discussed later
in the paper as well.


\subsection{Mesons in the presence of unquenched massive flavors}
\label{sec:KWmassive}

\begin{figure}[!ht]
\centering
\includegraphics[width=0.49\textwidth]{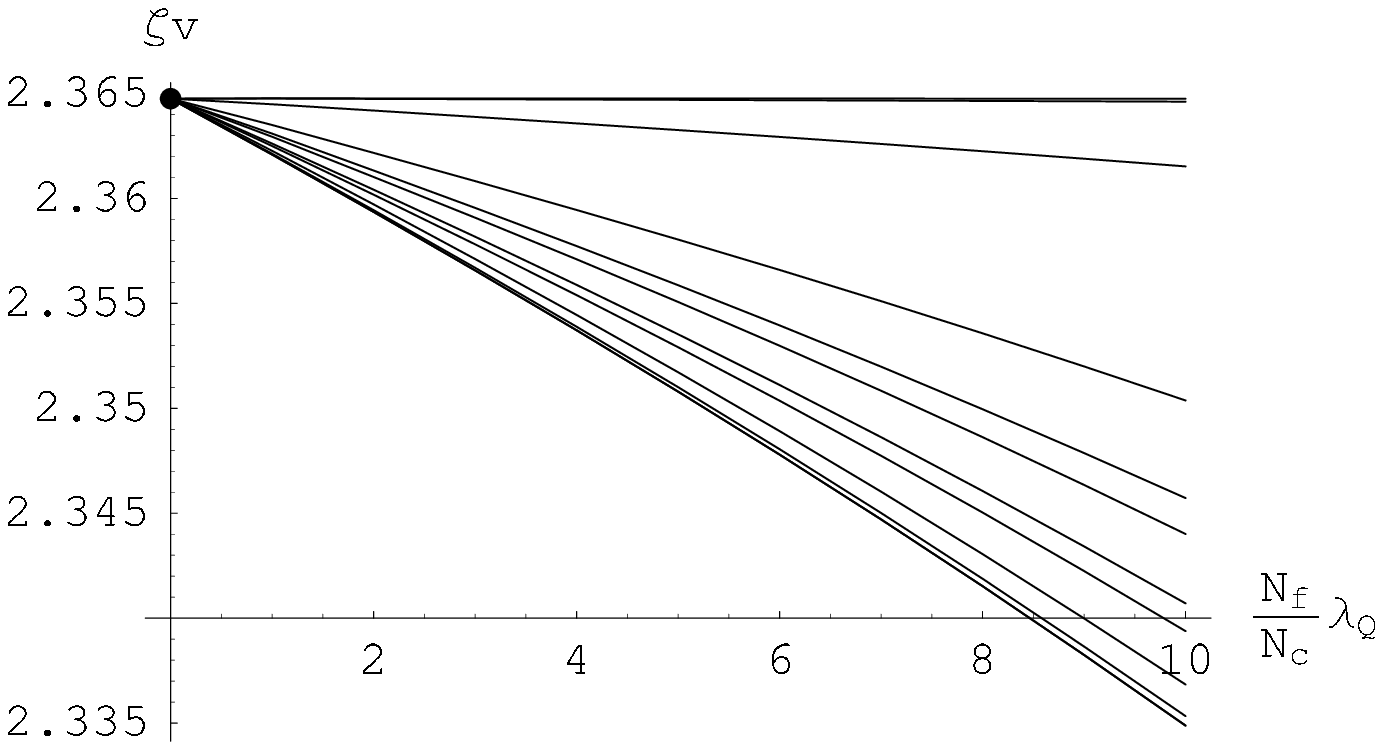} 
\includegraphics[width=0.49\textwidth]{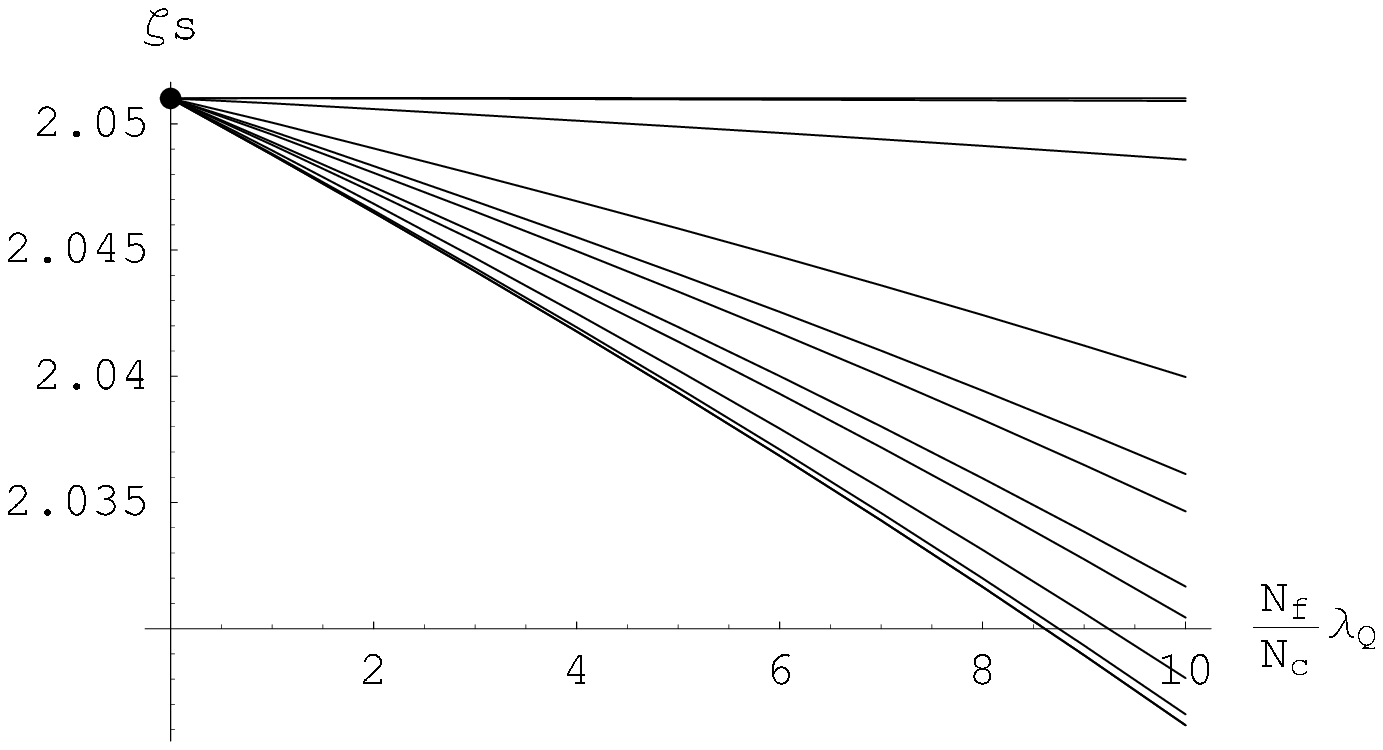} 
\caption{The functions $\z_{v,s}$ entering the meson mass formulae,
see eqs. (\ref{WKBKW}), (\ref{M-omega}). 
 In each plot, the different lines are for a series of values of
$\rho_q - \rho_Q$, namely $\rho_q - \rho_Q=-10,-5,-1,-0.5,-0.2,-0.1,
0.1,0.2,0.5,2,5,7$.}
\label{fig: massive1}
\end{figure}

We can generalize the analysis of the previous section to the case
in which the backreacting dynamical flavors are massive.
We take these dynamical flavors to be of the same (non-chiral) type
as the probe one, but with different mass $m_q$ (such that the tip
of the backreacting branes lies at $\rho_q$). The background geometry is obtained
by inserting in (\ref{firsordeqs}) the expression for $N_f(\rho)$
computed in \cite{Bigazzi:2008ie}, namely $N_f(\rho) = 0$ for
$\rho<\rho_q$ and $N_f(\rho) = N_f (1 - e^{3\r_q -3\r})$ 
for $\rho\geq \rho_q$. 
Let us define:
\bear
&& k_1 = 1+ 6(\rho_{LP}-\rho) +2 e^{3\rho_q} (e^{-3\rho_{LP}} - 2 e^{-3\rho})+
e^{6\rho_q -6\rho}\,\,,\quad\rc
&& k_2 =  6(\rho_{LP}-\rho) +2 e^{3\rho_q} (e^{-3\rho_{LP}} -  e^{-3\rho})\,\,,
\rc
&& k_q= 2 (e^{3\rho_q-3\rho_{LP}}-1) + 6 (\rho_{LP} - \rho_q)\,\,,
\eear
so the functions determining the background can be written as:
\bear
&&e^\phi = \frac{8\pi}{g_s N_f k_2}\,\,,\qquad e^g = c_3\,  k_1^{1/6}
e^{\rho - \rho_{LP}}\,\,,\qquad
e^f = c_3\frac{k_2^{1/2}}{k_1^{1/3}} e^{\rho - \rho_{LP}}\,\,,\rc
&&h = \frac{27\pi g_s N_c}{c_3^4} \int_\rho^{\rho_{LP}}  k_1^{-2/3}
e^{-4\rho +4 \rho_{LP}}d\rho\,\,,\qquad \qquad \qquad \qquad(\rho\geq\rho_q)\,\,,
\eear
and:
\bear
&&e^\phi = \frac{8\pi}{g_s N_f k_q}\,\,,\qquad\qquad\qquad e^g =e^f=c_3\,  k_q^{1/6}
e^{\rho - \rho_{LP}}\,\,,\rc
&&h = \frac{27\pi g_s N_c}{c_3^4}e^{4\rho_{LP}} \left( \int_{\rho_q}^{\rho_{LP}}  k_1^{-2/3}
e^{-4\rho }d\rho  +\frac14 k_q^{-2/3} (e^{-4\rho}-e^{-4\rho_q})\right)\,\,,
\,\qquad  (\rho\leq\rho_q)\,\,.\qquad\qquad
\eear
With this input, we can easily analyze the WKB integrals
(\ref{sigmaV}), (\ref{sigmaS}). Now, they depend on two parameters, namely
$\frac{N_f}{N_c}\lambda_Q$ and  the
mass of the dynamical quarks through the quantity
$\rho_q - \rho_Q$. Results are plotted in figure 
\ref{fig: massive1}.
At fixed $\rho_q - \rho_Q$, the meson masses slightly decrease with $\frac{N_f}{N_c}\lambda_Q$.
As $\frac{N_f}{N_c}\lambda_Q\to 0$, the
$\z$'s tend to the quenched values (\ref{WKBKWquen}).
Moreover, the larger the value of $\rho_q - \rho_Q$, 
the larger are the meson masses at fixed $\frac{N_f}{N_c}\lambda_Q$. 
For large $\rho_q - \rho_Q$, the
$\z$'s
tend to the quenched values too, {\it i.e.} the 
lines in figure 
\ref{fig: massive1} become horizontal.
This is expected since if the dynamical
quarks are very massive, they become quenched, even if there are many of them.
At small $\rho_q - \rho_Q\to -\infty$, one recovers the massless quark set-up of section
\ref{sec: masslessKW} and indeed the lower line in each plot is well approximated by (\ref{lamb}).

\subsection{On the constituent quark mass}
\label{sec: const}

As explained above, the definition (\ref{M-omega}),
which uses the local string tension, and (\ref{Mvs}), which
involves the constituent mass defined as in (\ref{quenchedmq}),
coincide in the quenched (conformal) case, but differ in the
non-conformal cases. We study here how the dimensionful
prefactors in these equations vary with respect to each other
when varying $\frac{N_f}{N_c}\lambda_Q$.  
Let us define the following quantity:
\be
\delta=\frac{\sqrt{2\pi} \left(\frac{32}{27}\right)^\frac14
 \frac{T_{Q}^\frac12}{\lambda_Q^\frac14}}
{\frac{2\pi m_Q}{\lambda_Q^\frac12}\ \sqrt{\frac{32}{27}}}\,\,.
\ee
By Taylor expanding for large $\bar\rho_{LP}$ (small $\frac{N_f}{N_c}\lambda_Q$),
we find the following leading order expression for $\delta$:
\bear
\delta&=&1 + \left[\left(\frac38 -\frac35 e^{\bar\rho_q}+\frac{3}{14} e^{3\bar\rho_q}\right)
\frac{3}{32\pi^2}\frac{N_f}{N_c}\lambda_Q\right] + \dots\qquad\qquad   \bar\rho_q <0\,\,,\rc
\delta&=& 1 - \left[\left(\frac{3}{280}e^{-4\bar\rho_q}\right)\frac{3}{32\pi^2}\frac{N_f}{N_c}\lambda_Q
\right]+ \dots\qquad\qquad \qquad\qquad  \bar\rho_q >0\,\,,
\eear
where we have defined $\bar\rho_q = \rho_q -\rho_Q$.
Both for $\frac{N_f}{N_c}\lambda_Q\to 0$
or $\bar\rho_q \to\infty$, one has $\delta\to 1$, recovering the
quenched result.
This relation, together with the results in figure  \ref{fig: massive1},
allows one  to compute how the $\omega$'s would vary with $\frac{N_f}{N_c}\lambda_Q$
if, instead of (\ref{M-omega})
 one decided to use (\ref{quenchedmq}), (\ref{Mvs}) as the definition of
$\omega$.

\subsection{The holographic $a$-function in flavored KW models}
\label{sec:holakw}
\begin{figure}
 \centering
\includegraphics[width=.4\textwidth]{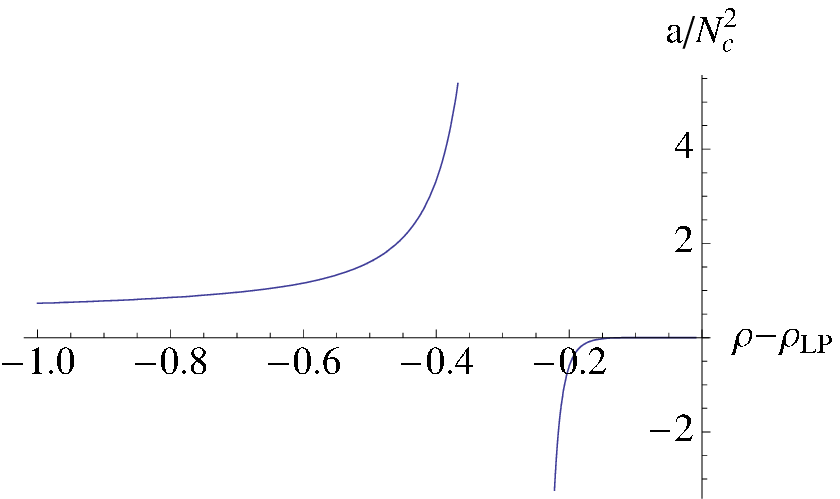} \hspace{1.5cm}
\includegraphics[width=.4\textwidth]{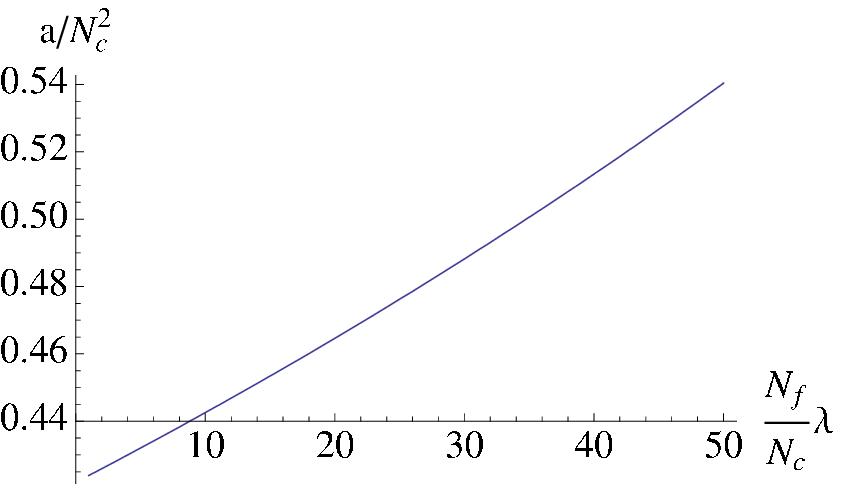}
\caption{The holographic $a$-function in the massless-flavored KW model as a function of $\rho-\rho_{Lp}$ (left) and of $(N_f/N_c)\lambda = (32\pi^2/3) (\rho_{LP}-\rho)^{-1}$ (right).}
\label{hola}
\end{figure}

As we have anticipated in section \ref{gene}, the D3-D7 conifold models under study are plagued by various UV pathologies. The behavior of the holographic $a$-function (see eq. (\ref{holadef})), in particular, suggests that any UV cutoff we choose to adopt in the analysis of the mesonic spectra has to be located below the discontinuity point $\rho_a$ where the function diverges. Fixing the overall constant factor such that $a=a_{KW}=N_c^2(27/64)$ in the unflavored case, the  function $a(\rho)$ for the flavored KW models can be taken as:
\be
a(\rho) = \frac{27}{8g_s^2\alpha'^{5/2}\pi^5}h^{3/2}\,e^{3f}\, H^{7/2} [H']^{-3}\,,
\label{holafkw}
\ee
where
\be
H(\rho)= \left(\frac{16\pi^3}{27}\right)^2\,h\,e^{2f+8g}{\alpha'}^5\,.
\ee
Let us now consider the flavored KW model with massless  quarks. 
The function $H(\rho)$,  which in the unflavored case is monotonically increasing as $e^{6\rho}$, now has  a maximum at $\rho_a\approx \rho_{LP}-0.27$.\footnote{$H(\rho)$
 starts from zero at $\rho\rightarrow-\infty$ and comes backs to zero at $\rho=\rho_{LP}$. This behavior does not depend on the particular choice of integration constant for the warp factor.} This behavior strongly affects that of the holographic $a$-function\footnote{It is simple to realize that $a(\rho)$ does not depend on $g_s$ and on the integration constant $c_3$. It just depends on $\rho_{LP}-\rho$ and is proportional to $N_c^2$.} as it 
is evident in figure \ref{hola}.
As expected, the holographic $a$-function has a bad discontinuity at $\rho_{a}$, where $H'(\rho_a)=0$.  For $\rho<\rho_{a}$, $a(\rho)$ is positive and increasing with $\rho$, whereas for $\rho>\rho_{a}$ it grows from minus infinity to zero. 

Far below the Landau pole, the holographic $a$-function is only slightly varying with $(N_f/N_c)\lambda$ and has an almost linear behavior
\be
a(\rho)\approx N_c^2\left[\frac{27}{64} + 2\times 10^{-3} (g_{FT}^2 N_f) \right] \qquad \quad(\rho\ll\rho_{LP})\,\,,
\ee
where $g^2_{FT}\sim e^{\phi(\rho)}$. This expression can be read as the leading correction to the effective number of adjoint plus bifundamental degrees of freedom, due to the internal quark loops (see analogous comments for the entropy of the D3-D7 model in flat space  in ref. \cite{termodbrane}).

\begin{figure}
 \centering
\includegraphics[width=.4\textwidth]{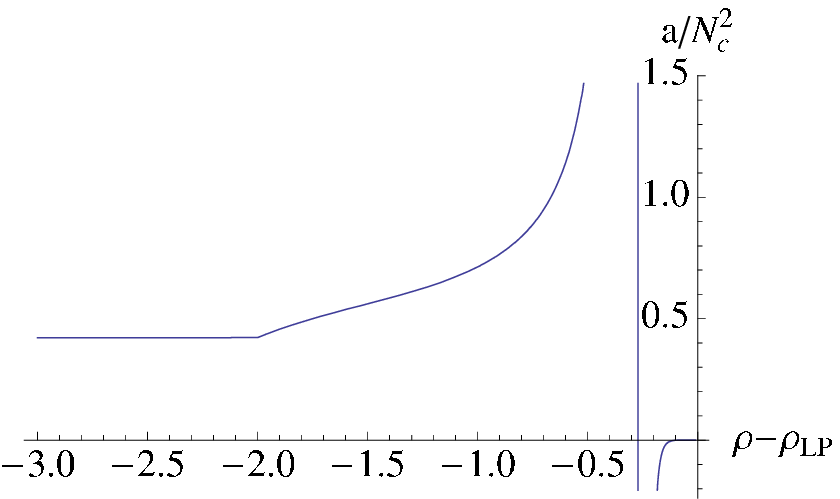} \hspace{1.5cm}
\includegraphics[width=.4\textwidth]{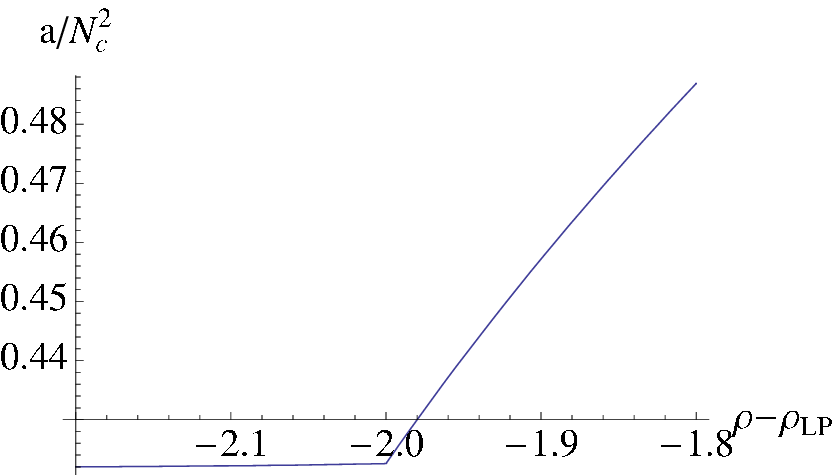}
\caption{The holographic $a$-function in the non chiral massive flavored KW model. We have fixed $\rho_q=\rho_{LP}-2$. On the right, a zoom around $\rho_q-\rho_{LP}$.}
\label{holamassivekw}
\end{figure}

For the non-chiral massive-flavored KW solution, where the sea quarks have a mass which is related to $\rho_q$, relevant plots are shown in figure \ref{holamassivekw}.
The holographic $a$-function approaches the unflavored value $a_{KW}$, for $\rho\ll\rho_q$; then it is very slowly increasing with $\rho$ up to $\rho=\rho_q$. Slightly  above $\rho_q$ the function $a(\rho)$ starts increasing faster and then it blows up towards a point $\rho_{a}$ where the previously noted bad discontinuity shows up. In the present case $\rho_a$ is a function of $\rho_q$: it goes to $\rho_{LP}-0.27$, as in the massless flavored case, for very small sea quark masses ($\rho_{LP}-\rho_q\rightarrow\infty$) and goes to $\rho_{LP}$ for very large masses.  
For $\rho>\rho_{a}$ the $a$-function is again going from minus infinity to zero. 

In section \ref{sec:FKWM}, we will suggest a possible relation between the behavior of the
$a$-function and that of the $\zeta_{v,s}$ functions in the mesonic spectrum.



\section{Meson excitations in the flavored KS models}
\label{sec: KS}
\setcounter{equation}{0}

This section follows the same structure as the previous one,
but we consider solutions in which the conifold metric is deformed and there
are three-form fluxes that affect the UV behavior of the background.
Namely, the quenched set-up is the Klebanov-Strassler solution
\cite{ks}. We discuss mesonic excitations in this background and in its
generalizations with unquenched massless \cite{Benini:2007gx}
and massive \cite{Bigazzi:2008qq} non-chiral flavors.
The solution with massless chiral flavors was found in \cite{benini},
but we will not address that case here.

\subsection{Background and excitation equations}

Let us write the metric in the notation of
\cite{kuper}.
In Einstein frame it reads (both for the quenched and
unquenched cases):
\bear
ds^2 &=& 
h^{-\frac12}dx_{1,3}^2 + h^{\frac12}\
\a' B^2(\tau) (d\tau^2 +(h_3 + \tilde h_3)^2) + \rc
&+& h^{\frac12}\ \a'A^2(\tau)
\left(h_1^2 + h_2^2 + \tilde h_1^2 + \tilde h_2^2
+ \frac{2}{\cosh\tau} (h_2 \tilde h_2 - h_1 \tilde h_1)\right)\,\,,
\label{deformedmetric2}
\eear
where we have used the definitions in (\ref{hdefs}).
There are non-trivial $F_{(1)}$, $F_{(3)}$, $F_{(5)}$, $H_{(3)}$ forms, which
we will not need in the following.
We refer the reader to \cite{Benini:2007gx} for details.

For the fluctuating probe brane, we will consider a supersymmetric embedding
first discussed in \cite{kuper}, where it was shown to correspond
to a massless non-chiral flavor in this set-up. Concretely, the embedding
is parameterized as $\gamma=0$, $\delta=const$, in terms of the variables
introduced in appendix \ref{technical-appendix}.
We discuss in the following mesonic excitations of this probe brane.
Note that this simple massless embedding cannot be used in the KW cases because it would not
produce a discrete spectrum, due to the different IR behavior of the backgrounds.
Oscillations around massive generalizations of this embedding 
(in the quenched background)
have been
discussed in \cite{Benini:2009ff}, but are beyond the scope of the
present work.

Similarly to (\ref{vectorans}), let us just consider excitations of the
gauge field of the form:
\be
{\cal A} =  a_v(\tau) e^{i kx} \xi_\m dx^\m+
a_s(\t) e^{i kx} h_1\,\,.
\ee
The corresponding second order equations are \cite{kuper}:\footnote{In
\cite{kuper}, a constant dilaton was considered but it turns out that a
running dilaton does not modify the expressions (\ref{KSmesoneqs}).
In the very far UV these equations coincide
with those of KW, as expected, see eqs. (\ref{aeq}), (\ref{beq}).
Looking at the far UV amounts to setting $e^{3\rho_Q-3\rho}=0$,
$\tanh \tau = \tanh \frac{\tau}{2}=\coth \frac{\tau}{2}=1$. One also
has to identify $e^{2g}=6A^2$, $e^{2f}=9B^2$ and $\tau=3\rho$.}
\bear
0&=&\partial_\tau\left(A^2  \tanh\tau
\partial_\tau a_v \right)+
M_v^2\, h\,\a' A^2 B^2 \tanh \tau\  a_v \,\,,\rc
0&=&\partial_\tau\left(\frac{\coth\frac{\tau}{2}}{h}
\partial_\tau a_s \right)+\left[
-\frac12 \partial_\tau (h^{-1})-
\frac14 \frac{\tanh\frac{\tau}{2}}{h}+
M_s^2\,\a'  B^2  \coth\frac{\tau}{2} \right] 
a_s \,\,.
\label{KSmesoneqs}
\eear
Again, we want to factor the meson masses as a dimensionful scale
related to certain IR quantities (evaluated at the tip of the
oscillating brane, namely $\tau=0$)
 times a purely numerical factor which
we will study. As a first step, let us introduce a similar factorization
for the function $h$:
\be
h=\left(e^{\phi}\right)\big|_{\tau=0}\
\frac{2^\frac23 (\alpha'g_sM)^2}{\mu^\frac83}\ I(\tau)\,\,,
\label{hhhdefi}
\ee
and let us define:
\be
M_{v,s}=
2\pi 2^\frac56 \frac{T_s^\frac12}{\lambda_0^\frac12}\ \omega_{v,s}\,\,,
\label{mvsks}
\ee
where we have introduced the string tension and the
't Hooft coupling\footnote{
As in footnote \ref{footcoupling}, we emphasize that we abuse of
language when calling $\lambda$ the `t Hooft coupling. } in the IR as:
\be
T_s = \frac{1}{2\pi\a'}\left(e^{\phi/2} \sqrt{G_{tt}G_{xx}}\right)|_{\tau=0}=
 \frac{1}{2\pi\a'^2} \frac{I_{0}^{-\frac12}\mu^\frac43}{2^\frac13 g_s M}\,\,,\qquad\qquad
\lambda_0 = 8\pi g_s M \left(e^{\phi}\right)\big|_{\tau=0}\,\,,
\label{Tslambdadef}
\ee
where we have defined $I_0 \equiv I|_{\tau=0}$.
Our goal in the following is to determine the value of the $\omega$'s 
in the set-ups with and without dynamical flavors.
As in the previous section, WKB estimates will be very useful.
 In this case, for both  the vector and scalar modes, the integral (\ref{Sigma-general}) which governs the behavior of the 
spectrum reads:
\be
\Sigma = \mu^{-\frac23} I_0^{-\frac14}  \int_0^{\tau_{cut}} I^\frac12 B\, d\tau\,\,,\qquad
\quad \zeta \equiv \frac{\pi}{\Sigma}\,\,.
\label{WKBKS}
\ee


\subsection{Mesons in the quenched approximation}

Let us write the value of the functions that determine the
geometry when no backreacting flavor branes are present
\cite{ks}. Define:
\be
K(\tau)=\frac{(\sinh (2\tau) - 2 \tau)^\frac13}{2^\frac13 \sinh\tau}\,\,.
\ee
Then, the different functions of the ansatz read:
\be
A^2(\tau)=\frac{\mu^\frac43}{4}\cosh \tau \,
K(\tau)\,\,,\qquad
B^2(\tau)=\frac{\mu^\frac43}{6K(\tau)^2}
\,\,,\qquad e^{\phi}=const\,\,,
\label{KSfunctions}
\ee
whereas the function $I$ is given in terms of the following integral:
\be
I(\tau) = I_{0}-
\int_0^\tau \frac{x\,\coth x-1}{\sinh^2 x}(\sinh(2x) - 2x)^\frac13 dx
\,\,.
\label{IKS}
\ee
The condition $I(\infty)=0$ fixes:
\be
I_{0}=0.71805\,\,.
\ee
In this case,  the second order equations (\ref{KSmesoneqs}) reduce to:
\bear
0&=&\partial_\tau\left(K(\tau) \sinh\tau
\partial_\tau a_v \right)+
\omega_v^2\, I_{0}^{-\frac12}I(\tau)\,\frac{\sinh\tau}{6K(\tau)}  \  a_v\,\,,\rc
0&=&\partial_\tau\left(\frac{\coth\frac{\tau}{2}}{I(\tau)}
\partial_\tau a_s \right)+\left[
-\frac12 \partial_\tau (I(\tau)^{-1})-
\frac14 \frac{\tanh\frac{\tau}{2}}{I(\tau)}+I_{0}^{-\frac12}
\omega_s^2\,\frac{1}{6K(\tau)^2}  \coth\frac{\tau}{2} \right] 
a_s \,\,.\qquad\qquad
\label{KSmesoneqsquen}
\eear
The discrete set of $\omega$'s which make the fluctuations regular in the
IR and normalizable in the UV can be determined numerically
via the shooting technique. We obtain:\footnote{
The same computation has been performed in \cite{kuper}.
Our numerical values differ from those displayed there
(even taking into account that there is a factor of
$I_{0}^{-\frac14}$ difference between our definition of $\omega$ and the
definition of $\lambda$ in \cite{kuper}).}
\bear
\omega_{v}&\approx & 1.367, 2.553, 3.752, 4.956, 6.164,\dots\rc
\omega_{s}&\approx & 3.100, 4.491, 5.778, 7.032, 8.271,\dots
\eear

\subsubsection{WKB estimates}

\begin{figure}[!ht]
\centering
\includegraphics[width=0.49\textwidth]{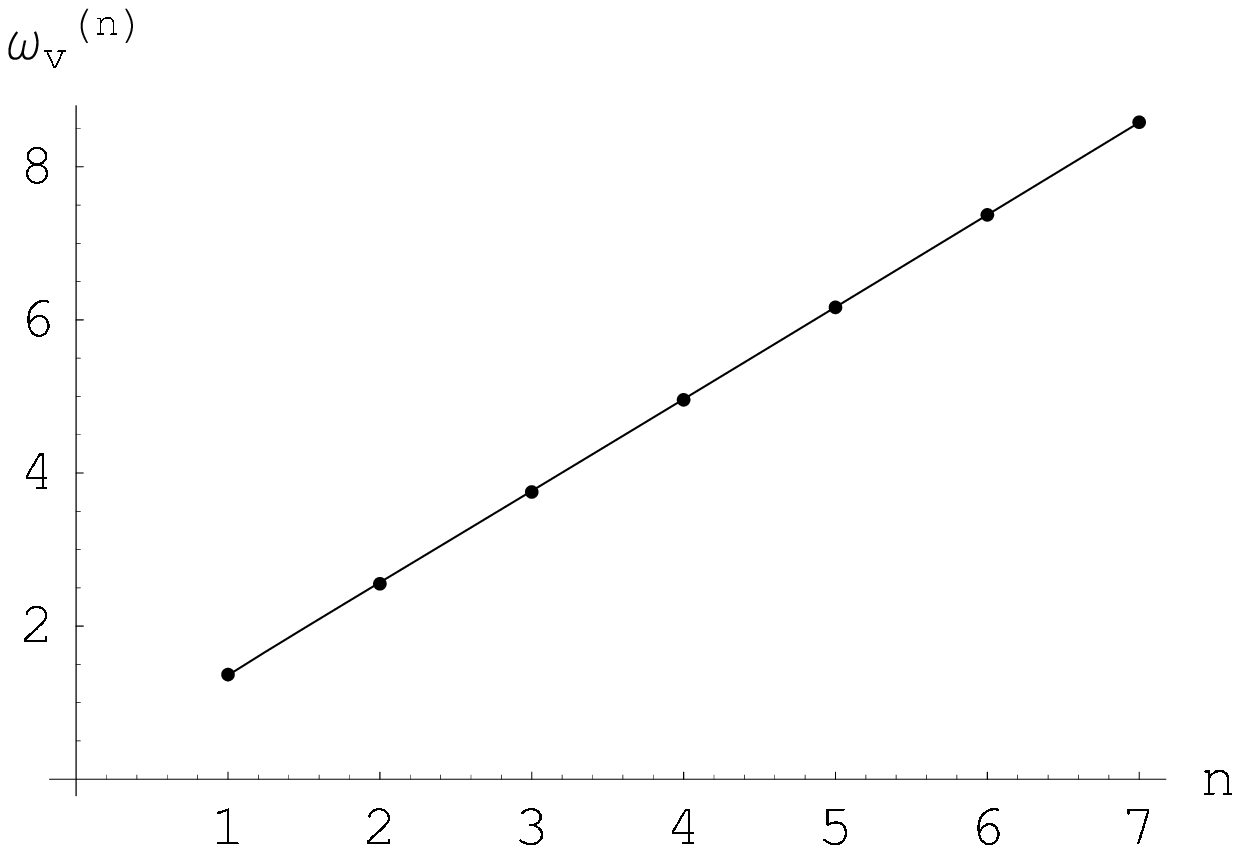} 
\includegraphics[width=0.49\textwidth]{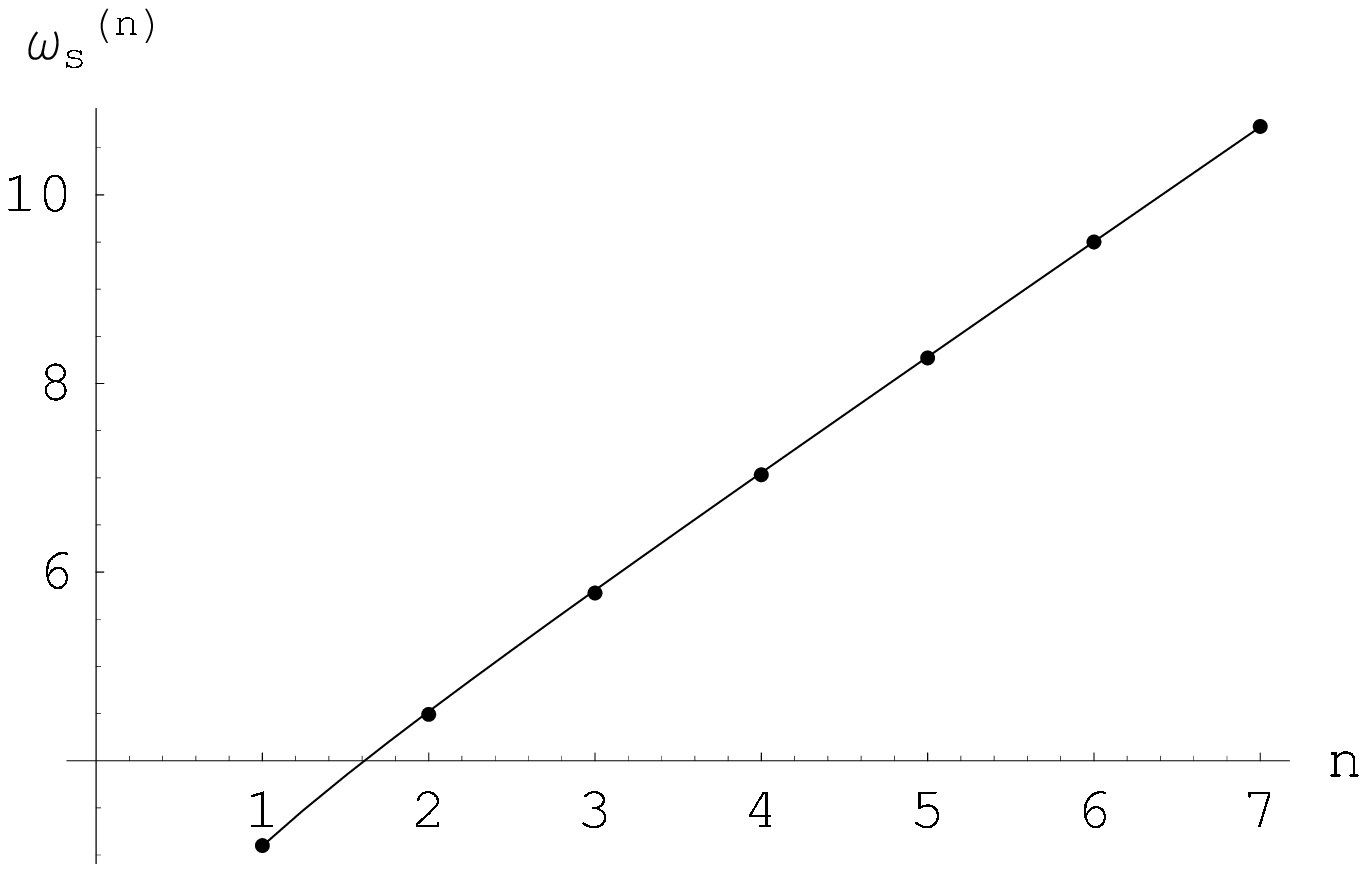} 
\caption{Comparison of the estimates (\ref{KSestimates}) to the first few points
obtained numerically. 
 }
\label{fig: KSmes}
\end{figure}

Applying the WKB approximation (as in appendix \ref{schr-appendix}) to this case,
we get the following estimates for the masses:
\bear
\omega_{v}&\approx &  \z \,\sqrt{n^2 +\frac13 \log n + \frac14}\,\,,\rc
\omega_{s}&\approx &  \z\, \sqrt{n^2 + \frac72 n + \frac54 \log n + 2}\,\,,
\label{KSestimates}
\eear
where $\z=1.214$ is what comes from the WKB integral (\ref{WKBKS}), in
which $\tau_{cut} = \infty$. 
The constant terms inside the square roots, as well as those with  logarithms, 
do not come from WKB but from fitting the points obtained numerically.
We have included those
terms in order to fit masses with low $n$ 
(the WKB result is only guaranteed to be accurate for large $n$). We stress 
that such dependences on $n$ are not exact, but just useful simple approximations.
A comparison of the expressions (\ref{KSestimates})
 to the first few numerical values is given in figure \ref{fig: KSmes}.

\subsection{Mesons in the massless unquenched set-up}

The deformation of the solution due to a set of smeared massless
flavors was found in \cite{Benini:2007gx}. 
In the following, we employ a  notation slightly different from the one used 
in \cite{Benini:2007gx}.
Let us call $\tau_d$ the position of the Landau pole and
define:
\be
\Lambda(\tau)= \frac{\left[2(\tau_d-\tau)(\sinh(2\tau) - \tau)
+ \cosh(2\tau) -2 \tau\,\tau_d -1\right]^\frac13}
{(4\tau_d)^\frac13 \sinh\tau}\,\,,
\ee
which is 
such that in the quenched limit $\lim_{\tau_d\to \infty} \Lambda(\tau)= K(\tau)$.
The other functions read:
\be
A^2(\tau) = \frac{\mu^\frac43}{4}\cosh \tau \,
\Lambda(\tau)\,\,,\qquad
B^2(\tau)=\frac{\mu^\frac43}{6\Lambda(\tau)^2}\,\,\frac{\tau_d-\tau}{\tau_d}
\,\,,\qquad
e^\phi = \frac{4\pi}{g_s N_f (\tau_d -\tau)}\,\,.
\label{BBBB}
\ee
The first two expressions reduce to those in (\ref{KSfunctions})
when $\tau_d\to \infty$. From the expression of the dilaton and the
definition of $\lambda_0$ in 
(\ref{Tslambdadef}), one gets:
\be
\tau_d = \frac{32\pi^2M}{N_f\lambda_0}\,\,.
\label{tau0Nf}
\ee
The remaining function entering the metric is 
$I(\tau)$ which can be written as:
\be
I(\tau)=I_0 - 2^{-\frac13} \tau_d^\frac53
\int_0^\tau \frac{x\,\coth x-1}{(\tau_d -x)^2\sinh^2 x}
\ \frac{-\cosh (2x) + 4 x^2 - 4x\,\tau_d +1-(x-2\tau_d)\sinh(2x)}
{(\cosh(2x) + 2x^2 -4x\,\tau_d -1 -2(x-\tau_d) \sinh(2x))^\frac23}dx
\label{itau}
\ee
which consistently 
reduces to  (\ref{IKS}) as $\tau_d \to \infty$.
We still have to give a condition that fixes $I_0$ for a 
given $\tau_d$. 
Notice that, regardless of  the value $I_0>0$ that we may
insert, the function $I(\tau)$ vanishes at some value 
$\tau_h < \tau_d$, since, in fact, it is easy to
check that $I(\tau_d)=-\infty$. 
However, when $\tau_d \gg 1$, there is a region 
$1\ll \tau <\tau_d - \epsilon$,  where the derivative of $I(\tau)$ is
exponentially small. The parameter
$\epsilon \sim \tau_d^\frac32 e^{-\frac23\tau_d}$
 is an exponentially small quantity defined by the fact that
the $(x-\tau_d)^{-2}$ in the integrand of (\ref{itau}) starts dominating
and makes $\partial_\tau I$ diverge.
We define $I_0$ as the  offset of $I(\tau)$ such that the 
almost flat region 
$1\ll \tau <\tau_d - \epsilon$ lies around $I=0$.
For instance, we can set $\tau_h=\tau_d-1$ and 
fix $I_0$ by 
requiring
$I(\tau_h)=0$. Any other similar definition, as for instance
$\tau_h=\tau_d-2$, would only affect $I_0$ by a negligible exponentially
small quantity (of order $e^{-\frac43\tau_h}$). 
Thus, as anticipated in section \ref{gene}, 
the choice of UV cutoffs 
only affects the IR physical quantities such as meson masses by negligible amounts.
We emphasize that for this reasoning to be valid, it is necessary that
$\tau_d \gg 1$, making it possible to have a large separation between the UV and
IR scales (see related comments in section \ref{sec: masslessSchro}).

Let us now write the excitation second order equations (\ref{KSmesoneqs})
for this case:
\bear
0&=&\partial_\tau\left(\Lambda(\tau) \sinh\tau
\partial_\tau a_v \right)+
\omega_v^2\, I_{0}^{-\frac12}I(\tau)\,\frac{\sinh\tau}{6\Lambda(\tau)} 
\frac{\tau_d -\tau}{\tau_d} \  a_v\,\,,
\label{KSmesoneqsquenm}
\\
0&=&\partial_\tau\left(\frac{\coth\frac{\tau}{2}}{I(\tau)}
\partial_\tau a_s \right)+\left[
-\frac12 \partial_\tau (I(\tau)^{-1})-
\frac14 \frac{\tanh\frac{\tau}{2}}{I(\tau)}+I_{0}^{-\frac12}
\omega_s^2\,\frac{\tau_d-\tau}{6\Lambda(\tau)^2\tau_d}  \coth\frac{\tau}{2} \right] 
a_s \,\,.\nonumber
\eear
The values of $\omega$ only depend on $\tau_d$ which, in turn, is related
through (\ref{tau0Nf}) to $\frac{N_f}{M}\lambda_0$, the physical parameter
controlling the deformation produced by the unquenched flavors.

Notice that the D3-D7 background in \cite{Benini:2007gx} has a (good) curvature singularity at $\tau=0$, the minimal distance reached by the D7-brane probe we are focusing on. Nevertheless, since the metric components are not singular there, we can extrapolate the fluctuations up to the origin where, as usual, we impose them to be regular.
\subsubsection{WKB estimates}

Similarly to section \ref{sec:KW}, the WKB formulae
(\ref{KSestimates}) also  match  the results in unquenched backgrounds quite well,
the main effect of the flavor backreaction being  a shift in the value
of $\zeta$. Inserting (\ref{BBBB}) in (\ref{WKBKS}) yields:
\be
\z = \pi  \left(\int_0^{\tau_{cut}} 
\frac{I_0^{-\frac14} I(\tau)^\frac12}{\sqrt6 \,\Lambda(\tau)}\frac{\sqrt{\tau_d - \tau}}{\sqrt{\tau_d}}
d\tau \right)^{-1}\,\,.
\label{zetaKS}
\ee
We will choose $\tau_{cut}<\tau_a \approx \tau_h - 0.38$. It
is worth stressing once again the independence of the result in such
UV prescription:
for  $1\ll \tau<\tau_h$, the
integrand is of order
$e^{-\tau/3}$ since
$I \propto e^{-\frac43 \tau}$ and  $\Lambda \propto e^{-\tau/3}$.
 Had we chosen a different $\tau_{cut}$, it would only have
changed the integral by an exponentially small quantity. In more
physical terms, the main contribution to the integral which fixes the 
meson masses comes from the IR region, as expected on general grounds.
Following the reasoning at the end of section \ref{sec:EMSW}, 
one can check that the size of the corrections
associated to the UV features is of order $\frac{\Lambda_{IR}}{\Lambda_{UV}}$.

By expanding for large $\tau_d$, we find:
\be\label{zetaksapprox}
\zeta \approx 1.214 ( 1 - 1.14 \times 10^{-2} \frac{N_f}{M}\lambda_0 + \dots)\,\,.
\ee


\subsection{Mesons in the presence of massive dynamical quarks}
\label{sec:massiveks}

We now consider mesons in the flavored KS solution with massive dynamical quarks \cite{Bigazzi:2008qq}.

Let us begin from the step function approximation for the function representing the effective number of flavor degrees of freedom, $N_f(\tau)=N_f\Theta[\tau-\tau_q]$, where $\tau_q$ is related to the mass of the dynamical quarks.  
The solution in the IR, for $\tau<\tau_q$, is just the ordinary KS solution but for the value of $I_0$, which is again set by the requirement that $h(\tau_{h})=0$.
For $\tau>\tau_q$ the solution is practically equal to the massless flavored one \cite{Benini:2007gx} considered in the previous section.
The only difference is in the integration constant, previously set to one, in the function $\Lambda$:
\be    
\Lambda(\tau)= \frac{\left[2(\tau_d-\tau)(\sinh(2\tau) - \tau)
+ \cosh(2\tau) -2 \tau\,\tau_d -(\cosh(2\tau_q)-2\tau_q^2)\right]^\frac13}
{(4\tau_d)^\frac13 \sinh\tau}\,\,,
\ee
and in a relation between the IR and UV scales $\mu$:
\be
\mu_{UV}^{4/3}=\mu_{IR}^{4/3} \Bigl(\frac{\tau_d}{\tau_d-\tau_q}\Bigr)^{1/3}.
\ee
These relations follow from the requirement of continuity of the metric at $\tau_q$.
Also, the coupling in the IR will now  read:
\be
\lambda_0 = 8\pi g_sM e^{\phi}|_{\tau=0}=\lambda_{\tau_q} = \frac{32\pi^2M}{N_f(\tau_d-\tau_q)}\,\,.
\label{tau0Nf2}
\ee
It does not run below the sea quark mass scale $\tau_q$.
Using the definition (\ref{hhhdefi})
we have, in the UV region:\footnote{Note that the continuity of the metric at $\tau_q$ implies a discontinuity of $I$:
$I_{UV}(\tau_q)=I_{IR}(\tau_q)\Bigl(\frac{\tau_d}{\tau_d-\tau_q}\Bigr)^{2/3}.$}
\bear
I(\tau)&=& - 2^{-\frac13}\tau_d^\frac23 (\tau_d-\tau_q)
\int^\tau \Bigl[ \frac{x\,\coth x-1}{(\tau_d -x)^2\sinh^2 x}
\ \cdot \nonumber \\
&& \frac{-\cosh (2x) + 4 x^2 - 4x\,\tau_d +1-(x-2\tau_d)\sinh(2x)}
{(\cosh(2x) + 2x^2 -4x\,\tau_d -(\cosh(2\tau_q)-2\tau_q^2) -2(x-\tau_d) \sinh(2x))^\frac23} \Bigr] dx\,\,.\qquad
\label{itau2}
\eear

The equations for the mesons are the same as in the previous sections and their mass is controlled in the WKB approximation by the function $\Sigma$.
It is calculated with the KS solution for $\tau<\tau_q$, while for $\tau>\tau_q$ it reads:
\be
\Sigma_{UV} = \int_{\tau_q}^{\tau_{cut}}\frac{I_0^{-\frac14} I(\tau)^\frac12}{\sqrt6 \,\Lambda(\tau)}\frac{\sqrt{\tau_d - \tau}}{\sqrt{\tau_d}}\Bigl(\frac{\tau_d-\tau_q}{\tau_d}\Bigr)^{1/6} \, d\tau \,\,.
\ee

\begin{figure}[!ht]
\centering
\includegraphics[width=0.6\textwidth]{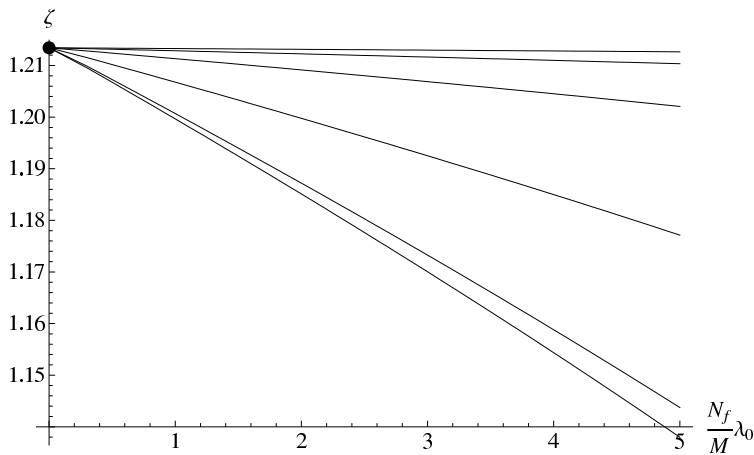} 
\caption{The plot of $\z$, see (\ref{WKBintKS2}), entering the 
WKB meson mass formulae. The different lines are, from bottom to top, for $\tau_q=0.1, 1, 5, 10, 15, 20$. 
The meson masses vary only very mildly when changing $\frac{N_f}{M}\lambda_0$. 
Also, they go to the quenched case both for small $\frac{N_f}{M}\lambda_0$ and for large quark mass.}
\label{KSfigmassive}
\end{figure}

The result for the function 
\be
\z  = \frac{\pi}{\Sigma}\,\,,
\label{WKBintKS2}
\ee
which gives the meson mass dependence on the number of flavors and their mass, 
is shown in figure \ref{KSfigmassive}.
The behavior of the masses is qualitatively 
the same as for the KW model, figure \ref{fig: massive1}.
The masses are mildly, almost linearly decreasing with $\frac{N_f}{M}\lambda_0$.
Moreover, the more massive  are the dynamical quarks, the smaller is the variation; for very large quark masses, the meson masses are practically constant and equal to the quenched values, as expected. 

We have repeated this calculation for the exact solution found in  \cite{Bigazzi:2008qq}, where the function $N_f(\tau)$ is not of the approximate step form but is explicitly calculated from the brane embeddings.
Since the formulae for this background are rather involved, we do not report them here and refer the reader to \cite{Bigazzi:2008qq}.
In any case, the qualitative behavior of the calculation of the meson masses is exactly as described above in the step approximation for  $N_f(\tau)$.
The plot of the function $\z$ is qualitatively identical to figure \ref{KSfigmassive}, also in the region not accessible by the step approximation, where the flavor branes reach the tip of the conifold and 
so the sea quark mass is smaller than the dynamically generated scale, but it is still non-zero.

\subsection{The holographic $a$-function}
\begin{figure}
 \centering
\includegraphics[width=.4\textwidth]{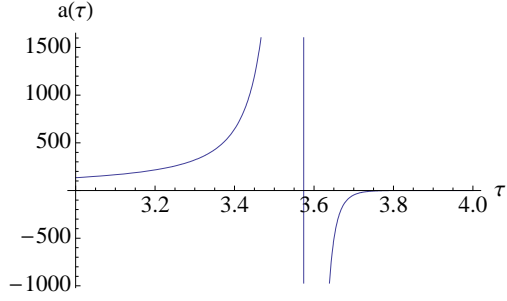}
\caption{The holographic $a$-function in the massless-flavored KS model. We have set $\tau_d=10, h(0)=1$ and fixed $2^{2/3}g_s(32\pi^2)M^2= 8\pi\mu^{8/3}N_f\tau_d$. In this case $\tau_h\approx 3.93$ and $\tau_a\approx 3.57$.}
\label{holaksnew}
\end{figure}

The expression for the holographic $a$-function in the flavored KS case is given by:
\bea
&&\beta(\tau)\equiv\alpha' \frac{\mu^{4/3}}{6}\frac{(\tau_d-\tau)}{\tau_d}\frac{h(\tau)}{\Lambda(\tau)^2}\,,\nonumber \\
&& H(\tau)\equiv \frac{8}{3}\pi^6{\alpha'}^5\mu^{20/3}h(\tau)\Lambda(\tau)^2\sinh^4(\tau) \frac{(\tau_d-\tau)}{\tau_d}\,,\nonumber \\
&& a(\tau) \sim 27 \beta^{3/2}H^{7/2}[H']^{-3}\,.
\eea
It is clear that the holographic $a$-charge can be real only for $\tau\le \tau_h$, where $h(\tau_h)=0$. For $\tau\rightarrow0$ the function goes to zero as
\be
a(\tau)\approx M^2\,\lambda_0^2\, I_0^2\, \tau^5\,,\quad (\tau\rightarrow0)\,,
\label{aksinzero}
\ee
and increases with $\tau$ up to the point $\tau_a<\tau_h$ (where $H'(\tau_a)=0$) where it diverges and it has the bad discontinuity discussed in section \ref{gene}. A representative plot of the holographic function $a(\tau)$ in the massless-flavored KS solution is given in figure \ref{holaksnew}.

\subsection{The domain wall tension}

We compute here an expression for the domain wall tension, which
is related to the value of the gluino condensate.
It comes from the action of a D5-brane sitting at $\tau=0$,
wrapping the finite $S^3$ \cite{Loewy:2001pq}, namely:
\be
T_{DW} = 16 \pi^2 T_5  \a'^{\frac32} \left(A^2 B\, e^{\frac{\phi}{2}}\, \right)\Big|_{\tau=0} =
\frac{1}{g_s} \, \frac{1}{2\pi^2 \a'^{3}} \left(\frac23\right)^\frac12\frac{\mu^2}{8}  \left( e^{\frac{\phi}{2}} \right)\Big|_{\tau=0}\,\,.
\ee
We can write it in terms of the IR quantities defined in
(\ref{Tslambdadef}):
\be
\frac{T_{DW}}{M\, T_s^\frac32 \,\lambda_0^\frac12} =  \frac{1}{8\pi^2 \sqrt3} I_0^\frac34\,\,.
\label{TDW}
\ee
This expression gives a  meaning to $I_0$ in terms of physical quantities. 
By expanding
for large $\tau_d$, one can find the effect of the 
(massless) unquenched quarks on this quantity
to be:
\be\label{izero}
I_0 \approx 0.71805 ( 1 + 1.16 \times 10^{-2} \frac{N_f}{M}\lambda_0 + \dots)\,\,.
\ee
\section{Towards an interpretation of the behavior of the mass spectra}
\label{rough}
\setcounter{equation}{0}
In the previous sections we have extracted the mass spectra for specific mesons in the KW and KS models with and without dynamical sea quarks, by studying fluctuations of the $U(1)$ gauge field on the worldvolume of a probe D7-brane.
The equations of motion for more generic fluctuations can be very difficult to solve and/or diagonalize and so
it could be useful to have an idea of how the mass spectra behave (as a function of the scales and couplings of the theory) in those cases. Moreover it would be nice to understand the behaviors we have found for the mesonic spectra at fixed scales or couplings in terms of simple general expressions.
\subsection{Unflavored KW-like models}
Let us start by reviewing some known results in the unflavored D3-brane models on the singular conifold, or,  in more generality, on a singular toric Calabi-Yau cone over a 5d Sasaki-Einstein compact manifold $X_5$. The gauge theory on the D3-branes has gauge group of the form $SU(N)^p$ where $p$ is an integer, and there are bifundamental and adjoint matter fields coupled through various superpotential terms. The dual supergravity background has a near horizon $AdS_5\times X_5$ metric: 
\be
ds^2 = \frac{r^2}{R^2} dx_{\mu} dx^{\mu} + \frac{R^2}{r^2} \left(dr^2 + r^2 ds_5^2\right)\,,
\ee
with $AdS$ radius  given by:
\be
R^4 = 4\pi g_s N_c \alpha'^2 \frac{\pi^3}{Vol(X_5)}\,.
\ee 
This background is dual to an IR surface of fixed points for the planar gauge theory at strong  (effective) 't Hooft coupling $\lambda=g_{FT}^2N_c\sim g_s N_c\gg1$, where $g_1=g_2=...=g_p\equiv g_{FT}$.  The volume of $X_5$ is holographically related to the IR central charge $a_{FT}$ as in  \cite{central}:
\be
a_{FT} = \frac{N_c^2}{4 Vol(X_5)} \pi^3\,.
\ee
Let us now place a (supersymmetric) probe D7-brane on $AdS_5\times X_5$, assuming that the brane reaches a minimal distance $r_Q$ from the origin of the space transverse to the D3-branes. The D3-D7 open strings will provide fundamental flavor fields of mass
$m_Q=r_Q/(2\pi\alpha')$. The low spin ($J=0,1$) mesons dual to fluctuating worldvolume fields on the D7-branes will have, just as in the simplest $X_5=S^5$ case \cite{kruc}, masses of the order:
\be
M\sim \frac{2\pi\alpha' m_Q}{R^2}\sim \frac{2 m_Q}{\sqrt{\lambda}}\sqrt{\frac{N_c^2}{a_{FT}}} \,,
\ee
where we omit universal (i.e. independent on $X_5$) numerical factors. 
Using the relation $2\pi\alpha' m_Q= r_Q = R\sqrt{G_{tt}(r_Q)}$ we can easily rewrite the formula above as:
\be
M\sim \frac{\sqrt{T_Q}}{\lambda^{1/4}}\left(\frac{N_c^2}{a_{FT}}\right)^{\frac14}\,,
\label{mkruc}
\ee
where $2\pi\alpha'T_Q\equiv G_{tt}(r_Q)$. This expression reproduces (in the constant $\lambda_Q=\lambda$ case) the parameterization of the meson masses introduced in eq. (\ref{M-omega}), once adapted to the conifold case, where $X_5=T^{1,1}$ with $Vol(T^{1,1})=16\pi^3/27$ and so $a_{FT}= N_c^2 (27/64)$.

As already outlined in the literature, these results imply that at strong coupling the binding energy of the $\bar Q Q$ system is 
very large, since the mesonic masses are parametrically smaller that the total quark mass $2m_Q$. Moreover, as was observed in \cite{strasslerq} for the $AdS_5\times S^5$ case, they imply that the typical size of the mesons is $L\sim \sqrt{\lambda}/m_Q$ which is very different from the weak coupling expression (the Bohr radius) $L_{weak}\sim (m_Q \lambda)^{-1}$ and much larger than $1/m_Q$.

For our purposes, a formula like (\ref{mkruc}) allows us to show how the meson masses
are expected to scale with the effective number of (adjoint+bifundamental) degrees of freedom (which are accounted for by $a_{FT}$). For fixed $T_Q/\sqrt{\lambda}$ the masses decrease as this number is increased. This behavior can be understood as an effect of the increasing binding between the constituent quarks as $a_{FT}$ is increased.

\subsection{Flavored KW models}
\label{sec:FKWM}

It is natural to think about a possible extension of formula (\ref{mkruc}) to the case where the probe D7-brane is put
in the fully backreacted D3-D7 background dual to, say, the flavored KW model. Since the fluctuating fields which are dual to the low spin mesons are localized on the probe D7-brane at $\rho_Q$ (we use, as in the rest of the paper, a different radial variable when referring to the unquenched solutions) we can expect the mesonic masses to scale as:
\be
M\sim \frac{\sqrt{T_Q}}{\lambda_Q^{1/4}}\left(\frac{N_c^2}{a_{Q}}\right)^{\frac14}\,,
\ee 
where $\lambda_Q\sim N_c e^{\phi(\rho_Q)}$ and $a_Q=a(\rho_Q)$ are the 't Hooft coupling and the holographic $a$-function (see eq. (\ref{holafkw})) evaluated at $\rho_Q$. The expression above could provide a simple interpretation of the linearly varying (with $N_f\lambda_Q/N_c$) functions $\zeta_{s,v}$ entering the mass parameterization (\ref{M-omega}):
\be
M_{s,v} \sim \frac{\sqrt{T_Q}}{\lambda_Q^{1/4}}\left(\frac{32}{27}\right)^{1/4}\zeta_{s,v}f_{s,v}(n)\,.
\label{prototip}
\ee
If we define:
\be\label{zetahere}
\zeta_Q \equiv \left(\frac{N_c^2}{a_Q}\right)^{\frac14}\,,
\ee
we can first notice that, just as $\zeta_{s,v}$, this function is indeed always decreasing (outside the pathological UV region) as $N_f\lambda_Q/N_c$ increases. In fact, as we have observed in section  \ref{sec:holakw}, the holographic $a$-function is always increasing slightly with $g_{FT}^2N_f\sim N_f\lambda/N_c$ as a result of the increasing number of effective (adjoint plus bifundamental) degrees of freedom due to the internal quark loops.

\begin{figure}
 \centering
\includegraphics[width=.4\textwidth]{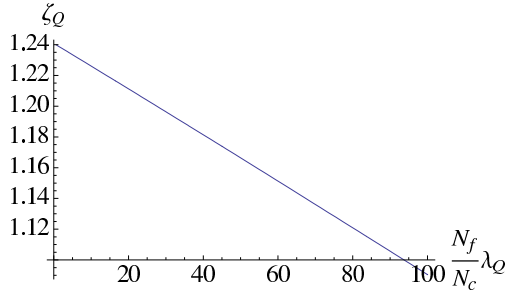}
\caption{The function $\zeta_Q$ (\ref{zetahere}) in the massless flavored KW model.}
\label{zetaq}
\end{figure}

Moreover, in the massless-flavored KW case, we can see that the function $\zeta_Q$  also has a quantitative behavior which is close to that of $\zeta_{s,v}$. Far below the Landau pole, $\zeta_Q$ is in fact a linearly decreasing function of $(N_f/N_c)\lambda_Q$ (see figure \ref{zetaq}).
The linear behavior persists up to large values ($O(200)$) of $(N_f/N_c)\lambda_Q$. The best linear fit is actually:
\be
\zeta_Q\approx \left(\frac{64}{27}\right)^{\frac14}\left(1 - 1.2 \times 10^{-3}\frac{N_f}{N_c}\lambda_Q\right)\,,
\ee
where the term in parentheses  is quite close to what we have found for $\zeta_{s,v}$ in eqns. (\ref{lamb}). 

Analogous comments apply to the massive-flavored KW model, though the quantitative matching between $\zeta_Q$ and $\zeta_{s,v}$ is realized only in the $\rho_Q>\rho_q$ region.

A proposal for an alternative way of interpreting the $\zeta_{v,s}$ behavior is briefly discussed in appendix \ref{lshift}.

\subsection{Flavored KS models}
A parameterization of the meson masses analogous to the one in (\ref{prototip}) has been adopted in section \ref{sec: KS} in the (flavored) KS cases. The only differences are that: 1) in those cases we have just focused on mesons made up of massless constituent quarks (the corresponding probe D7-brane reaching the origin); 2) the effective 't Hooft coupling was written in terms of the number $M$ of fractional D3-branes instead of that ($N_c$) counting the regular D3-branes. The first point just implies replacing $T_Q$ with $T_0=T_s$ and the second just means that we have to replace $\sqrt{\lambda_Q}$ with $\lambda_0$ in the mass formula. The resulting parameterization is that written in eq. (\ref{mvsks}). In order to give an interpretation to the $\zeta$ functions in the flavored KS cases one could thus think of naively extrapolating the previous results and so to relate those functions to $a_0^{-1/4}$, with $a_0$ being the holographic $a$-charge at the origin. However this reasoning would certainly fail, since $a_0=0$ in the KS models. 

For the cases where $m_Q$ is much larger than the IR dynamical scale, we expect the qualitative considerations proposed for the flavored KW models to apply.
Instead, we should consider alternative ways to interpret the $\zeta$ functions in the $m_Q=0$ flavored KS cases.
One interpretation is the following.
From the KS-like metrics we read that $T_s/m_{KK}^2 \sim \lambda_0 I_0^{1/2}$, where $m_{KK}$ is the mass scale of the glueball and KK modes.
Since in our analysis of the spectrum we keep fixed the ratio $(T_s/\lambda_0)^{1/2}$, we are keeping fixed $m_{KK}I_0^{1/4}$, which means
that $m_{KK}\sim I_0^{-1/4}$.
If we compare formulas (\ref{zetaksapprox}) and (\ref{izero}) we are led to argue that
 $\zeta\sim I_0^{-1/4}$ is a
measure of the variation of the KK (or glueball) mass scale with the
number of dynamical sea flavors and their masses. Possibly this variation can also be related to a change in the effective number of degrees of freedom around $\tau=0$. From eq. (\ref{aksinzero}), in fact, we get that $[a(\tau)/(M^2\tau^5)]^{-1/4}\sim \lambda_0^{-1/2} I_0^{-1/2}$ which displays a qualitative behavior similar to that related with $m_{KK}$.


\section {Summary and outlook}
\label{conclu}
\setcounter{equation}{0}
Since in real world QCD the number of flavors is of the same order of the number of colors, it is of obvious interest to study the extensions of the standard string/gauge theory correspondence to theories including the dynamical effects of fields in the fundamental representation \cite{localized,localized2,Mia:2009wj}.
Some technical difficulties in finding such extensions can be overcome if the branes carrying flavor degrees of freedom are smeared in the internal directions \cite{paris,tutti0}.
Despite the fact that, due to this simplification, there are nowadays a large number of available models in the literature \cite{Benini:2006hh}-\cite{Bigazzi:2008qq}, \cite{tutti0, tutti, tutti1}, some of their relevant features have not been fully investigated.
Among the others, an important issue concerns holography in the presence of a UV cutoff.  In this paper, we have addressed some relevant topics in the flavored versions of the Klebanov-Witten \cite{Benini:2006hh,Bigazzi:2008ie} and Klebanov-Strassler theories \cite{Benini:2007gx,Bigazzi:2008qq}, dual to D3-D7 brane systems on the (deformed) conifold.

We have started by refining the range of validity of the supergravity solutions.
Besides the point where the dilaton diverges, which is commonly associated to the presence of a Landau pole in the dual field theory, we uncovered in these backgrounds a point of singularity in the holographic $a$-function.
The existence of the singular point, which is located in the UV of the solution at least for small enough $N_f$, is clearly related to the presence of D7 branes.
\footnote{A possible explanation for such pathology is the breakdown of the naive interpretation of the gravity/field theory duality, due to the incomplete freezing of the gauge coupling on the D7 branes. In fact, if the D7 $SU(N_f)$ (or, better, $U(1)^{N_f}$) is not really a global symmetry but rather a (very) weakly gauged one, at a certain UV energy scale its gauge dynamics will be dominant over the D3 one.
At that point, the interpretation of the system as a 4d gauge theory coupled to $N_f$ flavors breaks down.
Thus, the latter interpretation is trustworthy only up to some UV cutoff scale.
An alternative explanation is that in the UV the claimed decoupling of the gauge theory with the bulk gravity modes is not fully realized. The presence of the singularity in the UV prevents us from casting any reasonable conclusion. We thank Carlos Nu\~nez and Francesco Benini for related comments.} The presence of the Landau pole already required a cutoff.
The new effect described here shifts the needed cutoff towards smaller energies.

Keeping in mind this issue, we have performed a study of small spin mesons in our conifold models.
The meson masses, even in the quenched case, are quite different from the QCD ones, due to the fact that the mesons are very tightly bound \cite{kruc}. 
So, there is no direct way of matching the corrections due to unquenching with actual QCD data. 
Nevertheless, our perspective is that the studies in the calculable gravity duals can be useful to gain some qualitative insights on the way (and to which extent) the screening effects can be understood from simple considerations on the physics of the models at hand.

Concretely, we have introduced a probe D7-brane in the backgrounds and have calculated the masses of some vector world-volume modes.
Our primary interest was to investigate how the meson masses depend on the dynamical flavor degrees of freedom, in particular their number and masses.
The presence of fields in the fundamental representation screens the charges, so the meson binding energy is affected by flavors.
The comparison of the screening effect among theories with different number and mass of the flavors requires some conventional choice of the physical scale to be kept fixed in the process.
Once this choice has been performed, the spectra of mesons have been calculated numerically, both for the quenched and the unquenched KW and KS theories.
Keeping fixed the coupling at the probe quark scale yields a decreasing of the meson masses as $g_{FT}^2 N_f$ is increased.
Had we kept fixed the coupling at some larger UV scale, larger $g_{FT}^2 N_f$ could have yielded larger meson masses.

Our computation of the change in the meson masses due to the unquenched
flavors is, in spirit, equivalent to the Lamb shift in atomic spectra: both
are effects that show up when taking into account the (slow) running of the
couplings in a theory with a Landau pole. The main qualitative difference is
that while in QED the theory is weakly coupled in the IR, in our case the
theory is always strongly coupled.

Finally, we have suggested that the mass dependence on the flavor parameters in the KW models seems to be simply described by the behavior of the holographic $a$-function, counting the number of effective degrees of freedom at a certain scale. In the KS cases it is more plausibly related to the changes in the
 glueball (or KK) mass scale as the dynamical flavor parameters are varied.

A relevant case that remained out of our investigation is the spectrum of mesons for massive probe quarks in the (flavored) KS model.
In general, the study of other observables in such flavored backgrounds could shed further light on the holographic correspondence in these models with UV cutoffs.
It would be interesting, in particular, to investigate the behavior of the entanglement entropy.
Finally, it would be important to understand holographic renormalization in theories with UV
cutoffs.

\vskip 15pt
\centerline{\bf Acknowledgments}
\vskip 10pt
\noindent
We are grateful to R. Argurio, F. Benini, C. Nu\~nez and J. Shock for relevant comments and to 
E. Imeroni and S. Kuperstein for useful discussions.
This work has been supported by the European Commission FP6 programme
MRTN-CT-2004 v-005104, ``Constituents, fundamental forces and symmetries in the universe''. F. B. is also supported  by the Belgian Fonds de la Recherche Fondamentale Collective (grant 2.4655.07), by the Belgian Institut Interuniversitaire des Sciences Nucl\'eaires (grant 4.4505.86) and
the Interuniversity Attraction Poles Programme (Belgian Science
Policy). A. C. is also supported by the FWO -
Vlaanderen, project G.0235.05 and by the Federal Office for
Scientific, Technical and Cultural Affairs through the Interuniversity Attraction
Poles Programme (Belgian Science Policy) P6/11-P. A. P. is also supported by a NWO VIDI grant 016.069.313 and by
INTAS contract 03-51-6346. A. R. is also supported by the MCINN and  FEDER (grant FPA2008-01838), the Spanish Consolider-Ingenio 2010 Programme CPAN (CSD2007-00042) and Xunta de Galicia (Conselleria de Educacion and grant PGIDIT06PXIB206185PR).

{\it F. B. and A. L. C. would like to thank the Italian students, parents and scientists for 
their activity in support of public education and research.}

\appendix

\setcounter{equation}{0}
\renewcommand{\theequation}{\Alph{section}.\arabic{equation}}

\section{Some technical details of section \ref{sec:KW}}
\label{technical-appendix}
\setcounter{equation}{0}
\subsection{The probe brane embedding}
\label{app: A1}

In order to study mesonic excitations, we consider a
 flavor D7-brane which preserves supersymmetry in the backgrounds (\ref{ansatzgeneral}). 
In particular, we take a probe brane embedding $z_4-z_3=\mu$
which introduces non-chiral flavors and was first studied
in detail by Kuperstein in \cite{kuper}. In the following,
we follow the steps of \cite{kuper} in order to introduce
a  new set of variables that will be convenient to describe
this embedding and its excitations.
The (dimensionless) $z's$ are defined as:
\bear
z_1 = e^{\frac32\rho}e^{\frac{i}{2}(\psi-\varphi_1-\varphi_2)}
\sin\frac{\theta_1}{2}\sin\frac{\theta_2}{2}\,,\qquad\quad
z_2 = e^{\frac32\rho}e^{\frac{i}{2}(\psi+\varphi_1+\varphi_2)}
\cos\frac{\theta_1}{2}\cos\frac{\theta_2}{2}\,,\rc
z_3 = e^{\frac32\rho}e^{\frac{i}{2}(\psi+\varphi_1-\varphi_2)}
\cos\frac{\theta_1}{2}\sin\frac{\theta_2}{2}\,,\qquad\quad
z_4 = e^{\frac32\rho}e^{\frac{i}{2}(\psi-\varphi_1+\varphi_2)}
\sin\frac{\theta_1}{2}\cos\frac{\theta_2}{2}\,.
\label{zangles}
\eear
The conifold equation $z_1z_2=z_3z_4$ can be written as:
\be
\det W=0\,\,,\qquad {\rm with}\qquad W=
\left(
\begin{array}{cc}
-z_3 & z_2 \\
-z_1 & z_4 
\end{array}
\right)\,\,.
\label{Wdef}
\ee
In order to write it in terms of the angles, let us introduce
$SU(2)$ matrices:
\be
L_i=
\left(
\begin{array}{cc}
\cos\frac{\theta_i}{2}e^{\frac{i}{2}(\psi_i+\varphi_i)} & 
- \sin\frac{\theta_i}{2}e^{-\frac{i}{2}(\psi_i-\varphi_i)}\\
\sin\frac{\theta_i}{2}e^{\frac{i}{2}(\psi_i-\varphi_i)} & 
\cos\frac{\theta_i}{2}e^{-\frac{i}{2}(\psi_i+\varphi_i)}
\end{array}
\right)\,\,.
\ee
Since the $\psi_i$ only appear in the combination $\psi_1 + \psi_2$,
let us set $\psi_1 = \psi_2 = \frac12 \psi$. We can explicitly 
check from (\ref{zangles}), (\ref{Wdef}) that:
\be
W = e^{\frac32\rho}\,L_1 \left( \begin{array}{cc}
0&1\\ 0& 0 \end{array} \right) L_2^{\dagger}\,\,.
\ee
 Now, the trick of \cite{kuper} consists of  introducing a change of coordinates
which eliminates $\theta_2, \varphi_2$ in
favor of two new angles $\gamma$, $\delta$ with ranges 
$\gamma\in [ 0,\pi], \delta \in [0,4\pi)$,
as:
\be
L_2 = L_1 S\,,\qquad {\rm with}
\qquad
S=\left(
\begin{array}{cc}
\cos\frac{\gamma}{2} e^{i\frac{\delta}{2}} & - i \sin
\frac{\gamma}{2} e^{-i\frac{\delta}{2}} \\
- i \sin
\frac{\gamma}{2} e^{i\frac{\delta}{2}} & 
\cos\frac{\gamma}{2} e^{-i\frac{\delta}{2}}
\end{array} \right)\,\,.
\label{L1L2S}
\ee
The nice point is that this new system of coordinates is very
well adapted to the massless non-chiral embedding $\tr W=
z_4-z_3=\mu$, since by explicit computation we find that:
\be
\tr W = i \,\sin\frac{\gamma}{2} e^{i\frac{\delta}{2}}
e^{\frac32\rho} = \mu\,\,.
\ee
Suppose without loss of generality
 that $\mu$ is real and positive so we can define
$\mu\equiv e^{\frac32 \rho_Q}$. Then, the $z_4-z_3=\mu$
embedding is just:
\be
\sin\frac{\gamma}{2}e^{\frac32\rho}=e^{\frac32\rho_Q}\,\,,\qquad
\delta=3\pi\,\,.
\label{embeq}
\ee
This embedding equation is valid both for the
quenched and unquenched set-ups. 
In particular, in the unquenched case, it is possible  
to show that (\ref{embeq})
gives a solution for the worldvolume action
$S_{D7} = T_7 \int( - e^\phi \sqrt{-\det(P[G_{\mu\nu}])} + P[C_{(8)}] )$,
where the $P$ denotes the pull-back on the brane worldvolume.
To do this, it is  necessary to use (\ref{firsordeqs})
and take into account that $C_{(8)}$ 
(defined as
$dC_{(8)} = e^{2\phi} ({}^* F_{(9)})$) takes a non-trivial
value in the background:
\be
C_{(8)} = - \frac{g_s N_f}{4\pi}\,\frac{e^{2\phi+4g}}{24}\,
\sin \theta_1 \cos \gamma \left(dx_0 \wedge
dx_1 \wedge dx_2 \wedge dx_3 \wedge
 d\r \wedge d\theta_1 \wedge d\varphi_1
\wedge d\psi\right)\,\,.\qquad
\ee

\subsection{The meson excitation equations}
\label{app: A2}

In order to study the fluctuations, we want to write down
the metric in terms of the new coordinates $\gamma,\delta$. 
Let us start by
rewriting the original metric in terms
of two sets of left invariant one forms:
\bear
h_1 &=& -\cos\frac{\psi}{2} \sin \theta_1 d\varphi_1
+\sin\frac{\psi}{2} d\theta_1\,\,,\qquad
h_2 = -\sin\frac{\psi}{2} \sin \theta_1 d\varphi_1
-\cos\frac{\psi}{2} d\theta_1\,\,,\rc
\tilde h_1 &=& -\cos\frac{\psi}{2} \sin \theta_2 d\varphi_2
+\sin\frac{\psi}{2} d\theta_2\,\,,\qquad
\tilde h_2 = -\sin\frac{\psi}{2} \sin \theta_2 d\varphi_2
-\cos\frac{\psi}{2} d\theta_2\,\,,\rc
h_3&=&\frac{d\psi}{2} + \cos\theta_1 d\varphi_1\,\,,\qquad\qquad
\qquad\qquad
\tilde h_3=\frac{d\psi}{2} + \cos\theta_2 d\varphi_2\,\,.
\label{hdefs}
\eear
The angular part of the metric in (\ref{ansatzgeneral}) reads:
\be
ds_5^2 = \frac{e^{2g}}{6} (h_1^2 + h_2^2 + \tilde h_1^2 + \tilde h_2^2)
+\frac{e^{2f}}{9} (h_3 + \tilde h_3)^2\,\,.
\label{t11met}
\ee
Substituting the expression above we get the usual metric written in terms
of the angles $\theta_1, \varphi_1, \theta_2, \varphi_2, \psi$.
The form of the $\tilde h_i$'s in terms of 
$h_i,\gamma,\delta$ can be obtained by explicit computation from (\ref{L1L2S})
\cite{kuper}:
\bear
\tilde h_1 &=& (h_1 - d\gamma) \cos \delta - (h_3 \sin\gamma
+ h_2 \cos \gamma) \sin \delta\,\,,\rc
\tilde h_2 &=& (h_1 -d\gamma)\sin\delta + (h_3 \sin\gamma
+ h_2 \cos \gamma) \cos \delta\,\,,\rc
\tilde h_3 &=& (h_3 \cos \gamma -h_2 \sin \gamma) + d\delta\,\,.
\label{kuperdefs}
\eear
Substituting in (\ref{t11met}) we get:
\bear
ds_5^2 &=& \frac{e^{2g}}{6} \left(\frac12 d\gamma^2 + 2\left(h_1-\frac{d\gamma}{2}\right)^2
+ h_2^2 + (\sin \gamma h_3 + \cos \gamma h_2)^2\right)+\rc
&&+\frac{e^{2f}}{9} (d\delta +h_3(1+\cos \gamma) - h_2 \sin\gamma)^2\,\,.
\label{t11metnew}
\eear
The induced worldvolume metric is obtained by inserting (\ref{embeq}) in
this expression.
In order to compute the  meson spectrum for the modes 
discussed, we insert
the ansatz for the fluctuation (\ref{vectorans}) in
the equations of motion stemming from 
the worldvolume action:
\bear
S_{D7} &=& T_7  \Bigl(-\int d^8x e^\phi\sqrt{ \det(-P[G_{\mu\nu}] + 2\pi\alpha'
e^{-\frac{\phi}{2}} F_{\mu\nu})}\nonumber \\
&& \quad \quad + \int P[C_{(8)}] + \frac{(2\pi\alpha')^2}{2} \int P[C_{(4)}]\wedge F \wedge F
\Bigr) \,\,.
\eear
Keeping only linear terms in the fluctuation equation,
 one finds (\ref{aeq}), (\ref{beq}).

\section{Schr\"odinger form and WKB estimates}
\label{schr-appendix}
\setcounter{equation}{0}
In this appendix we shall start by reviewing the mapping of a generic mesonic  fluctuation equation to a Schr\"odinger wave equation and the subsequent use of the WKB approximation to get an estimate of the corresponding mass levels.  We
shall follow closely section 3 of  ref. \cite{RS}, which we will adapt to our notations. 
Let us suppose that the fluctuation is represented by a function
$\phi(\bar\rho)$, which satisfies a differential equation  of the form:
\beq
e^{-\bar\rho}\,\partial_{\bar\rho}\,\big(\,e^{-\bar\rho}\,F(\bar\rho)\,\partial_{\bar\rho}\,\phi\,\big)\,+\,
\big(\, \omega^2\,H(\bar\rho)\,+\,P(\bar\rho)\,\big)\,\phi\,=\,0\,\,,
\label{ODE}
\eeq
where  the radial variable $\bar\rho$ is non-negative ($\bar \rho\ge 0$),  $ \omega$ is a mass parameter and $F(\bar\rho)$, $H(\bar\rho)$ and  $P(\bar\rho)$ are three arbitrary functions that are independent of $ \omega$.  We will assume that $\bar\rho$ takes values in the range $0\le\bar\rho\le \bar\rho_{cut}$, where $\bar\rho_{cut}$ is large and, in the quenched cases, we will just take  $\bar\rho_{cut}=+\infty$.

By performing a suitable  change of variables, eq. (\ref{ODE})  can be written as the zero energy Schr\"odinger equation:
\beq
\partial_y^2\,\psi\,-\,V (y)\,\psi\,=\,0\,\,,
\label{Schrodinger}
\eeq
where $V (y)$ is some potential to be determined. The change of 
variables needed to pass from  (\ref{ODE}) to (\ref{Schrodinger}) is \cite{RS}:
\beq
e^{y}\,=\,e^{\bar\rho}-1\,\,,\qquad\qquad
\phi\,=\,\frac{e^{\frac{y}{2}}}{\sqrt{F}}\,\psi\,\,.
\label{change-y}
\eeq
Notice that $y\ge - \infty$. 
In order to write the expression for the potential $V(y)$, 
let us define:
\beq
F_0\,\equiv\,e^{-y}\,F\,\,,\qquad
H_0\,\equiv\,e^{y}\,H\,\,,\qquad
P_0\,\equiv\,e^{y}\,P\,\,.
\eeq
Then,  the Schr\"odinger  potential  $V$  becomes \cite{RS}:
\beq
V\,=\,\frac12\,\frac{F_0''}{F_0}\,-\,\frac14\,\frac{F_0'^2}{F_0^2}\,-\,\frac{P_0}{F_0}\,-\,
\omega^2\,\frac{H_0}{F_0}\,\,,
\label{Sch-pot}
\eeq
where the primes denote derivatives with respect to $y$.  Notice that the potential $V$ in (\ref{Sch-pot}) depends parametrically on $\omega$.  When $V$ satisfies some general conditions, $\omega$ can be fine-tuned to some particular discrete set of values in order to produce a normalizable zero-energy solution of the Schr\"odinger equation (\ref{Schrodinger}).

One can get rather accurate estimates of the mass levels by means of the WKB approximation, whose starting point is the standard semiclassical quantization rule for the Schr\"odinger equation, namely:
\beq
(n-\frac12)\pi\,=\,\int_{y_1}^{y_2}\,dy\,\sqrt{-V(y)}\,\,,
\,\,\,\,\,\,\,\,\,\,\,\,\,\,
n\ge 1\,\,,
\label{WKBquant}
\eeq
where $n\in\ZZ$ and  $y_1$ and $y_2$ are the turning points of the 
potential 
($V(y_1)=V(y_2)=0$).  One can evaluate the right-hand 
side of eq. (\ref{WKBquant})
by expanding it as a power series in $1/ \omega$. By keeping the 
leading and subleading terms of
this expansion, one can obtain $\omega$ as a function of the quantum 
number $n$. Actually, as shown in \cite{RS}, one can obtain the WKB levels by simply looking at the IR ($\bar\rho\to 0$) and UV ($\bar\rho\to\infty$) behavior of the functions $F$, $H$ and $P$ entering the original fluctuation equation (\ref{ODE}).  Let us assume that near $\bar\rho \approx 0,\infty$ these functions behave as:
\bear
&&F\approx F_1\bar \rho^{s_1}\,\,,
\,\,\,\,\,\,\,\,\,\,\,\,\,\,
H\approx H_1 \bar\rho^{s_2}\,\,,
\,\,\,\,\,\,\,\,\,\,\,\,\,\,
P\approx P_1 \bar\rho^{s_3}\,\,,
\,\,\,\,\,\,\,\,\,\,\,\,\,\,{\rm as}\,\,\bar\rho\to 0\,\,,\rc\rc
&&F\approx F_2\,e^{r_1\bar\rho}\,\,,
\,\,\,\,\,\,\,\,\,\,\,\,\,\,
H\approx H_2\, e^{r_2\bar\rho}\,\,,
\,\,\,\,\,\,\,\,\,\,\,\,\,\,
P\approx P_2\, e^{r_3\bar\rho}\,\,,
\,\,\,\,\,\,\,\,\,\,\,\,\,\,{\rm as}\,\,\bar\rho\to \infty\,\,,
\label{asymp-FHP}
\eear
where $F_i$, $H_i$, $P_i$, $s_i$ and $r_i$ are constants.  The consistency of the WKB approximation requires \cite{RS} that 
$s_2-s_1+2$ and $r_1-r_2-2$ are
strictly positive numbers, whereas $s_3-s_1+2$ and $r_1-r_3-2$ can be 
either positive or
zero.   Let us define, 
following ref. \cite{RS},
the quantities $\alpha_1$ and $\beta_1$ as:
\beq
\alpha_1\,=\,s_2-s_1+2\,\,,
\,\,\,\,\,\,\,\,\,\,\,\,\,\,
\beta_1\,=\,r_1-r_2-2\,\,,
\label{alpha1-beta1}
\eeq
and $\alpha_2$ as:
\beq
\alpha_2\,=\,\begin{cases}
|s_1-1|& {\rm if}\,\,  s_3-s_1+2\not =0\,\, ,\cr\cr
\sqrt{(s_1-1)^2\,-\,\,\frac{4P_1}{F_1}} & {\rm if}\,\,  s_3-s_1+2 =0\,\, .
\end{cases}
\label{alpha2}
\eeq
Similarly, we define $\beta_2$ in the form:
\beq
\beta_2\,=\,\begin{cases}
|r_1-1|& {\rm if}\,\,  r_1-r_3-2\not =0\,\, ,\cr\cr
\sqrt{(r_1-1)^2\,-\,\,{4P_2\over F_2}} & {\rm if}\,\,  r_1-r_3-2 =0\,\, .
\end{cases}
\label{beta2}
\eeq
Notice that $\alpha_{1,2}$ are determined by the IR behavior of the fluctuation equation, whereas $\beta_{1,2}$ can be extracted from the UV limit of $F$, $H$ and $G$. 
Actually, the mass levels for large quantum number $n$ can be written in terms 
of $\alpha_{1,2}$ and $\beta_{1,2}$ as \cite{RS}:
\beq
 \omega^2_{WKB}\,=\,{\pi^2\over 
\Sigma^2}\,n\,\bigg(n\,-1\,+\,{\alpha_2\over \alpha_1}\,+\,
{\beta_2\over \beta_1}\bigg)\,\,,\quad\quad (n\ge 1)\,\,,
\label{generallWKB}
\eeq
where $\Sigma$ is the following integral:
\beq
\Sigma\,=\,\int_{0}^{\bar\rho_{cut}} d\bar\rho \,\,e^{\bar\rho}\,\,\sqrt{{H(\bar\rho)\over F(\bar\rho)}}\,\,.
\label{Sigma-general}
\eeq

\subsection{WKB for KW backgrounds}

Let us now particularize our formalism to the general case of a metric 
parameterized
by two functions $f$ and $g$ and a warp factor $h$ as the one written in (\ref{ansatzgeneral}). The differential equations for the vector and scalar modes (\ref{vectorans}) have been written in eqs. (\ref{aeq}) and (\ref{beq}) respectively. Let us rewrite these equations in terms of the variable  $\bar\rho=\rho-\rho_Q$ defined in (\ref{bardef}). One has:
\bear
&&\partial_{\bar \rho}\left(e^{2g-3\bar\rho}(e^{3\bar\rho}-1)\partial_{\bar \r} a_v \right)
+ \omega_v^2  \,\, {e^{2g+2f} \,I\over \sqrt{I_Q}} 
\left(1+ e^{- 3\bar\r} (\frac34 e^{2g-2f} -1) \right)\, a_v\,=\,0\,\,,
\label{aeq-barrho}\\\rc
&&\partial_{\bar \rho}\left(\frac{e^{3\bar\r}-1}{e^{3\bar\r}\,I}\partial_{\bar\r} a_s \right)\,+\,
\Bigg[\,{\omega_s^2\over \sqrt{I_Q}}\,\Bigg(\,e^{2f}\,+\,
e^{-3\bar\rho}\,\Big(\,{3\over 2}\,e^{2g}\,-\,e^{2f}\,\Big)\,\Bigg)\,-\,\rc\rc
&&\qquad\qquad\qquad\qquad\qquad\qquad
-{3\over 2}\,\partial_{\bar\rho}\big(I^{-1}\big)\,-\,{9\over 4}\,{e^{3\bar\rho}\over 
(e^{3\bar\rho}\,-\,1\,)\,I}\,\,\Bigg]\,a_s\,=\,0\,\,,
\label{beq-barrho}
\eear
where $f$, $g$  should be considered as functions of $\bar\rho$,  $\omega_{v,s}$ have been defined in  (\ref{M-omega}) and $I$ and $I_Q$ are given by:
\beq
I(\rho)\,\equiv\,{4\over 27\pi g_s N_c}\,h(\rho)\,\,,\qquad\qquad
I_Q\,\equiv\,I(\rho=\rho_Q)\,\,.
\eeq
By comparing the fluctuation equation (\ref{aeq-barrho}) with our general formula (\ref{ODE}), 
it follows  that the integral $\Sigma_v$ determining the mass gap for the vector modes   is:
\beq
\Sigma_{v}\,=\,\int_{0}^{\bar\rho_{cut}}\,
\Bigg[{e^{2f(\bar\rho)+3\bar\rho}\,+\,{3\over 4}\,e^{2g(\bar\rho)}\,-\,e^{2f(\bar\rho)}\over
e^{3\bar\rho}\,-\,1}\,\,{I(\bar\rho)\over \sqrt{I_Q}}\,
\Bigg]^{{1\over 2}}\,\,.
\label{sigmaV}
\eeq
Similarly,  for the scalar modes in (\ref{beq-barrho})  the corresponding $\Sigma$-integral is:
\beq
\Sigma_{s}\,=\,\int_{0}^{\bar\rho_{cut}}\,
\Bigg[{e^{2f(\bar\rho)+3\bar\rho}\,+\,{3\over 2}\,e^{2g(\bar\rho)}\,-\,e^{2f(\bar\rho)}\over
e^{3\bar\rho}\,-\,1}\,\,{I(\bar\rho)\over \sqrt{I_Q}}\,
\Bigg]^{{1\over 2}}\,\,.
\label{sigmaS}
\eeq

\subsubsection{Vector modes in the quenched KW}

\begin{figure}[!ht]
\centering
\includegraphics[width=0.4\textwidth]{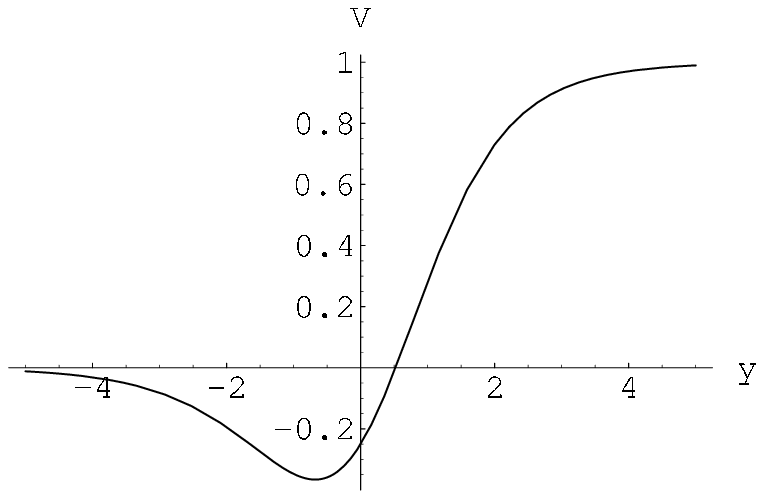} 
\qquad\qquad
\includegraphics[width=0.4\textwidth]{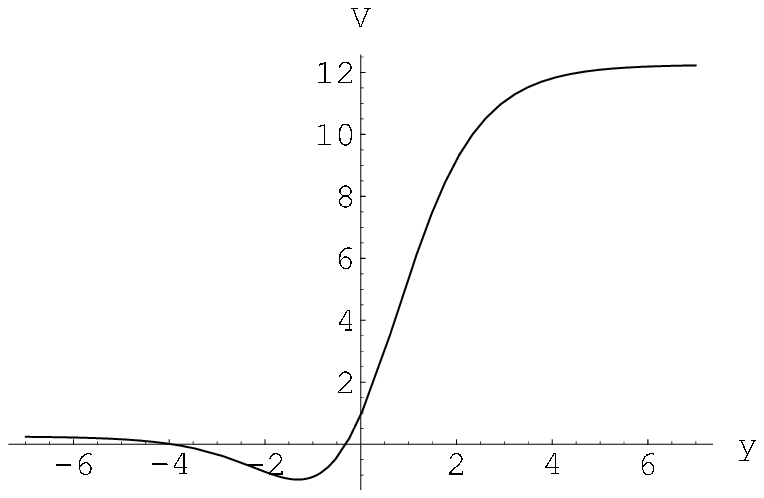} 
\caption{Schr\"odinger potentials for the
vector (left) and scalar (right)  excitations in the quenched  KW model. We used $\omega_v=3$ and $\omega_s=6$.}
\label{VschroKW}
\end{figure}

Let us apply our previous formalism to the case in which the background is just the KW without including the backreaction of the flavor branes. In this case we must simply take $f=g=\bar\rho+\rho_Q$ and $I=e^{-4\bar\rho-4\rho_Q} $ in our general formulas (see (\ref{unflavoredKW})). One can check that, in this case, the fluctuation equations (\ref{aeq-barrho}) and (\ref{beq-barrho}) reduce to the ones written in (\ref{nonchiveceq}) and (\ref{nonchira1eq}). 
Notice that in this case the range of $\bar\rho$ is just $0\le\bar\rho\le\ +\infty$. After removing  a common constant factor the  functions $F$, $H$ and $P$ take the form:
\beq
F(\bar \rho)\,=\,e^{3\bar\rho}\,-\,1\,\,,\qquad\qquad
H(\bar\rho)\,=\,e^{-\bar\rho}\,\Big(\,1\,-{e^{-3\bar\rho}\over 4 }\,\Big)\,\,,\qquad\qquad
P(\bar\rho)=0\,\,.
\label{FHP-quenchedKW-v}
\eeq
Changing variables as in (\ref{change-y}) we can get the expression of the Schr\"odinger potential (\ref{Sch-pot}) for this case. One gets an expression of the type:
\beq
V_v(y)\,=\,{e^y\over 4 (e^y+1)^4\,(e^{2y}\,+\,3 \,e^y\,+\,3 )^2}\,\,{\cal P}_v(y)\,\,,
\eeq
where ${\cal P}_v(y)$ is a polynomial in $e^y$ of degree seven, namely:
\bear
&&{\cal P}_v(y)\,=\,4\,e^{7y}\,+\,34\,e^{6y}\,+\,(129-4\omega_v^2)\,e^{5y}\,+\,
(274-24 \omega_v)\,e^{4y}\,+\,(346-60\omega_v^2) e^{3y}\,+\,\rc\rc
&&\qquad\qquad\qquad
+\,(258-75\omega_v^2) e^{2y}\,+\,15 (7-3 \omega_v^2) e^{y}\,+\,9\,(2-\omega_v^2)\,\,.
\eear
It follows from the above expression that:
\beq
\lim_{y\to-\infty}\,V_v(y)\,=\,0\,\,,\qquad\qquad
\lim_{y\to\infty}\,V_v(y)\,=\,1\,\,.
\eeq
In general, when $ \omega_v$ is above some threshold  (related to the meson mass gap), the potential $V$ has a minimum around $y=0$, where $V_v$ is negative  (see figure \ref{VschroKW}). 

Let us now estimate the WKB energy levels for the quenched vector modes. By analyzing the behavior of the functions (\ref{FHP-quenchedKW-v}) near $\bar\rho\approx 0$, we get that the coefficients defined in (\ref{asymp-FHP}) are:
\beq
F_1\,=\,3 \,\,,\qquad\qquad
s_1\,=\,1\,\,,\qquad\qquad
H_1\,=\,{3\over 4 }\,\,,\qquad\qquad
s_2\,=\,0\,\,,
\eeq
while $P_1=s_3=0$. Moreover, the behavior at $\bar\rho\to\infty$ is characterized by:
\beq
F_2\,=\,1\,\,,\qquad\qquad
r_1\,=\,3\,\,,\qquad\qquad
H_2\,=\,1\,\,,\qquad\qquad
r_2\,=\,-1\,\,,
\eeq
with $p_2=r_3=0$. Using these values in eqs. (\ref{alpha1-beta1})-(\ref{beta2}) we get:
\beq
\alpha_1\,=\,2\,\,,\qquad\qquad
\alpha_2\,=\,0\,\,,\qquad\qquad
\beta_1\,=\,2\,\,,\qquad\qquad
\beta_2\,=\,2\,\,,
\eeq
and then, as $\alpha_2/\alpha_1\,+\,\beta_2/\beta_1\,=\,1$, the WKB formula for the masses becomes:
\beq
\omega_{v}\,=\,{\pi\over \Sigma_{v}}\,\,n\,\,,
\label{omegav-general}
\eeq
with:
\beq
\Sigma_{v}\,=\,{1\over 2}\,\,\int_{0}^{\infty}\,d\bar \rho\,e^{-\bar\rho}{\sqrt{4e^{3\bar\rho}-1}\over 
\sqrt{e^{3\bar\rho}-1}}\,=\,\,{\sqrt{\pi}\,\Gamma\Big({4\over 3}\Big)\over 
\Gamma\Big({5\over 6}\Big)}\,F\Big(-{1\over 2},{1\over 3}, {5\over 6}, {1\over 4}\Big)\,\,.
\label{xi-v-unquenched}
\eeq
Numerically $\Sigma_{v}\,\approx\,1.3285$, which, after substituting this value in (\ref{omegav-general}), yields the expression of $\omega_{v}$ displayed in (\ref{WKBKW}).

\subsubsection{Scalar modes in the quenched KW}
Similarly, by using the functions $f$, $g$ and $h$ of (\ref{unflavoredKW}) in (\ref{beq-barrho}),  we can study the scalar fluctuations in the quenched KW case. We get:
\beq
F(\bar\rho)\,=\,e^{2\bar \rho}\, \,(e^{3\bar \rho}\, -\,1)\,\,,\qquad
H(\bar\rho)\,=\,{2 e^{3\bar \rho}\, +\,1\over 2 e^{2\bar \rho}}
\,\,,\qquad
P(\bar\rho)\,=\,-6e^{3\bar \rho}\,\Big[\,1\,+\,{3\over 8}\,
{e^{3\bar \rho}\over e^{3\bar \rho}-1}\,\Big]\,\,,
\eeq
where, again,  we have removed  an irrelevant constant factor. 
The potential $V_s$ associated to these functions takes the form:
\beq
V_s(y)\,=\,{1\over 4 (e^y+1)^4\,(e^{2y}\,+\,3 \,e^y\,+\,3 )}\,\,{\cal P}_s(y)\,\,,
\eeq
where now the polynomial ${\cal P}_s(y)$ is:
\bear
&& {\cal P}_s(y)\,=\,49\,e^{6y}\,+\,247\, e^{5y}\,+\,(501-4\omega_s^2)\, e^{4y}\,+\,
 (514-12 \omega_s^2)\, e^{3y}\,+\,\rc
 &&\qquad\qquad\qquad\qquad
 +\, (271-12\omega_s^2)\, e^{2y}\,+\,(63-6\omega_s^2)\, e^{y}\,+\,3\,\,.
\eear
Then:
\beq
\lim_{y\to-\infty}\,V_s(y)\,=\,{1\over 4}\,\,,\qquad\qquad
\lim_{y\to\infty}\,V_s(y)\,=\,{49\over 4}\,\,.
\eeq

In figure \ref{VschroKW} we have plotted $V_s$ for some typical value of $\omega_s$. As before, when $\omega_s$ is above the mass gap, $V_s$ has a unique minimum, where the potential is negative.

In order the get the WKB estimate of the mass levels, let us look at the behavior of the functions $F$, $H$ and $P$ at $\bar\rho=0$. We get:
\beq
F_1\,=\,3 \,\,,\qquad
s_1=1\,\,,\qquad
H_1\,=\,{3\over 2} \,\,,\qquad
s_2\,=\,0\,\,,\qquad
P_1\,=\,-{3\over 4} \,\,,\qquad
s_3\,=\,-1\,\,,
\eeq
whereas, by looking at $\rho\to\infty$ one concludes that:
\beq
F_2\,=\,1\,\,,\qquad
r_1=5\,\,,\qquad
H_2\,=\,1\,\,,\qquad
r_2\,=\,1\,\,, \qquad
P_2\,=\,-{33\over 4} \,\,,\qquad
r_3\,=\,3\,\,.
\eeq
Using these values, one arrives at the following coefficients:
\beq
\alpha_1\,=\,1\,\,,\qquad
\alpha_2\,=\,1\,\,,\qquad
\beta_1\,=\,2\,\,,\qquad
\beta_2\,=\,7\,\,.
\eeq
Notice that $\beta_2$ must be computed from the square root in (\ref{beta2}), which gives the integer value $7$. Moreover, in this case:
\beq
\Sigma_s\,=\,{1\over \sqrt{2}}\, \int_{0}^{\infty}\,d\bar\rho\, e^{-\bar\rho}\,
{\sqrt{2 e^{3\bar\rho}+1}\over 
\sqrt{e^{3\bar\rho}-1}}\,=\,\,{\sqrt{\pi}\,\Gamma\Big({4\over 3}\Big)\over 
\Gamma\Big({5\over 6}\Big)}\,F\Big(-{1\over 2},{1\over 3}, {5\over 6}, -{1\over 2}\Big)\,\,.
\eeq
Numerically $\Sigma_s\,\approx\,1.53173$ and the WKB mass formula becomes:
\beq
\omega_{s}\,=\,2.051\,\sqrt{n\Big(n+{7\over 2}\Big)}\,\,.
\eeq
To improve the agreement of this WKB estimate with the numerical results for small $n$ we can  add a new term independent of $n$ inside the square root in such a way that 
 the numerical values are fitted. It turns out that an excellent fit is obtained when this new term is $15/8$. The resulting expression of  $\omega_{s}$ is just the one written in (\ref{WKBKW}). 
\section{An alternative way to interpret the $\zeta$'s}
\label{lshift}
\setcounter{equation}{0}
Let us comment on a possible alternative way to interpret the $\zeta$ functions entering the expression for the mesonic masses. Let us write the mesonic mass formula in the following way:
\be
M\sim \frac{\sqrt{T(\mu)}}{\lambda^{\frac14}(\mu)}\,,
\label{guess}
\ee
where, taking inspiration from the Lamb shift calculations in QED, we take the energy scale $\mu$ to be related to the (modulus of the) binding energy $E_b$ of the meson, defined by $M = 2m_Q - E_b$. Let us take $\mu = |E_b|/2$ for example. Since $M=\epsilon\, 2m_Q$, with $\epsilon\ll1$ related to the inverse 't Hooft coupling (at the scale $m_Q$), we have:
\be
E_b\approx 2m_Q (1-\epsilon) \,.
\ee
This relation suggests to relate the energy scale $\mu$ with a radial variable:
\be
\rho\approx \rho_Q (1 - \epsilon_Q)\,,
\ee
and so to get, by a simple expansion of (\ref{guess}), a formal relation like:
\be
M\sim \frac{\sqrt{T_Q}}{\lambda_Q^{\frac14}}\left[1 + \epsilon w(\rho_Q)\right]\,,
\ee
where the function $w(\rho_Q)$ has to be determined case by case.
The above term in parentheses could possibly provide an alternative interpretation of the $\zeta$ functions. We leave this issue for future studies. 
\section{Comments on the quark-antiquark potential and high spin mesons}
\label{highspin}
\setcounter{equation}{0}
Let us consider the holographic expression for the potential $V(L)$ felt by a $\bar Q Q$ pair probing our D3-D7 models. Obviously, due to the presence of UV cutoffs in our set-ups, we cannot take these sources as being strictly static: their mass $m_Q$, in any case, will have to be smaller than the UV scale. 

The $\bar Q Q$ meson is dual to an open string attached to the probe D7-brane (at $\rho_Q$), bending in the bulk up to a minimal radial position $\rho_{tip}$. The Minkowski separation $L$ between the test quarks, as well as the potential $V(L)$ (i.e. the total energy renormalized by subtraction of the static quark masses) depend on $\rho_{tip}$. For an open string embedding given by $t=\tau,y=\sigma, \rho=\rho(y)$ where $y\in [-L/2,L/2]$ is one of the
spatial Minkowski directions, one finds  \cite{maldawilson}:
 \bear L(\rho_{tip})&=&2
\int_{\rho_{tip}}^{\rho_{Q}} \frac{G P_{tip}}{P\sqrt{P^2
-P_{tip}^2}}d\rho\,,\rc 
V (\rho_{tip})&=&\frac{2}{2\pi\alpha'} \Bigl[\int_{\rho_{tip}}^{\rho_{Q}}
\frac{G P}{\sqrt{P^2 -P_{tip}^2}}d\rho -
\int_{-\infty}^{\rho_{Q}} G \ d\rho \Bigr]\,, 
\label{maldafor}
\eear
where $P, G$ are expressed in terms of the string frame metric (we do not consider BH backgrounds so $G_{tt}=G_{yy}$):
\be
P = G_{tt} \,,\quad\qquad G = \sqrt{G_{tt}G_{\rho\rho}}\,,
\label{sonnendefs}
\ee 
and the ``$tip$'' subindex means that the quantity is evaluated at $\rho=\rho_{tip}$. From the relations above it is clear that the short distance behavior of $V(L)$ is not Coulomb-like. Instead, at $L=0$, i.e. at $\rho_{tip}=\rho_Q$, we get $V(0)=-2m_Q$, where:
\be
m_Q \equiv \frac{1}{2\pi\alpha'}\int_{-\infty}^{\rho_{Q}} G \, d\rho\,.
\ee  
Expanding the previous expressions around $L=0$ one actually gets:
\be
V(L) = -2m_Q + T_Q L + ...\qquad \left(L \ll \frac{2m_Q}{T_Q}\right)\,\,,
\ee
i.e. a linear potential, where the effective string tension $T_Q$ is given by $2\pi\alpha' T_Q = G_{tt}(\rho_Q)$. This result is generic and only depends on the fact that we are considering a D7-brane at finite $\rho_Q$: indeed it was previously noticed for a probe brane on $AdS_5\times S^5$ in \cite{kruc}. 
 
The occurrence of this linear potential at small $L$ explains why in \cite{kruc} the masses of mesons with spin $1\ll J\ll \sqrt{\lambda}$ follow linear Regge trajectories. The very same behavior is thus expected for the analogous mesons in the (flavored) KW models. The Regge slope dependence on the sea flavor parameters will be simply accounted for by the expression for $T_Q$. The study of the screening effects on this slope will thus easily follow. 

Mesons with spin $J\gg \sqrt{\lambda}$ should behave instead as non-relativistic quark-antiquark pairs (also at strong coupling) interacting through the large $L$ limit of $V(L)$ which, in the KW-like set-ups, is Coulomb-like. The mesonic masses in the flavored KW cases in such regime should have expressions analogous to those in \cite{kruc} and the study of screening effects on these mesons should not be a difficult  task.


\end{document}